\documentclass[journal]{IEEEtran}
\usepackage{pdfpages}
\usepackage{amsmath}
\usepackage{cite}
\usepackage{mathtools}

\usepackage{multicol}
\usepackage{soul}
\ifCLASSINFOpdf
\usepackage{graphicx}
\usepackage{multirow}
\usepackage{txfonts}
\usepackage{dblfloatfix}
\else
\usepackage{graphicx}
\fi
\usepackage{float}
\ifCLASSOPTIONcompsoc
\usepackage[caption=false,font=footnotesize,labelfont=sf,textfont=sf]{subfig}
\else
\usepackage[caption=false,font=footnotesize]{subfig}
\fi
\usepackage{cite}
\usepackage{dblfloatfix}
\usepackage{algorithm}
\usepackage{algpseudocode}
\usepackage{breqn}
\DeclareMathOperator*{\argmax}{arg\,max}
\usepackage{float}

\usepackage{caption}

\usepackage{lineno,hyperref}
 \usepackage{float}
\usepackage{multicol}
\usepackage{lipsum}
\usepackage{array}
\usepackage[acronym]{glossaries}
\usepackage{hhline}
\usepackage{url}
\usepackage{dblfloatfix}
\usepackage{lipsum}
\usepackage{fixltx2e}
\usepackage{tabu}

\usepackage{hyperref}
\hypersetup{
    colorlinks=true,
    linkcolor=black,
    filecolor=black,      
    urlcolor=black,
}
 
\begin{document}

\definecolor{amber(sae/ece)}{rgb}{1.0, 0.49, 0.0}
\definecolor{aqua}{rgb}{0.0, 1.0, 1.0}

\title{Towards Intelligent Reconfigurable Wireless Physical Layer (PHY)}


\author{Neelam Singh*, S. V. Sai Santosh*, and Sumit J. Darak
	\thanks{*Neelam Singh and S. V. Sai Santosh are joint first authors.}
					\thanks{This work is supported by the funding received from DST INSPIRE and core research grant (CRG) awarded to Dr. Sumit J. Darak from DST-SERB, GoI.}
			\thanks{Neelam Singh, S. V. Sai Santosh, and Sumit J. Darak are with Electronics and Communications Department, 
				IIIT-Delhi, India-110020 (e-mail: \{neelami,siripurapu17197,sumit\}@iiitd.ac.in)}
}
	\maketitle
\begin{abstract}
Next-generation wireless networks are getting significant attention because they promise 10-factor enhancement in mobile broadband along with the potential to enable new heterogeneous services. Services include massive machine type communications desired for Industrial 4.0 along with ultra-reliable low latency services for remote healthcare and vehicular communications. \textcolor{black}{In this paper, we present the design of \textit{intelligent} and \textit{reconfigurable} physical layer (PHY) to bring these services to reality. First, we design and implement the reconfigurable PHY via a hardware-software co-design approach on system-on-chip consisting of the ARM processor and field-programmable gate array (FPGA). The reconfigurable PHY is then made intelligent by augmenting it with online machine learning (OML) based decision-making algorithm. Such PHY can learn the environment (for example, wireless channel) and dynamically adapt the transceivers' configuration (i.e., modulation scheme, word-length) and select the wireless channel on-the-fly. Since the environment is unknown and changes with time, we make the OML architecture reconfigurable to enable dynamic switch between various OML algorithms on-the-fly. We have demonstrated the functional correctness of the proposed architecture for different environments and word-lengths. The detailed throughput, latency, and complexity analysis validate the feasibility and importance of the proposed intelligent and reconfigurable PHY in next-generation networks.}


\end{abstract}
\begin{IEEEkeywords}
 Intelligent reconfigurable architecture, Zynq SoC, physical layer, machine learning, multi-Armed Bandit
 \end{IEEEkeywords}

\section{Introduction}
Upcoming 5G and 6G networks are expected to enable smarter agriculture, remote health-care access, Industry 4.0, and intelligent transportation. To bring this to reality, they are envisioned to support enhanced mobile broadband (high throughput), mission-critical services (reliability and ultra-low latency), and a massive Internet of Things (lower data rates, ultra-dense deployment, and extended battery life) \cite{2017_NR_parkvall,2018_NR_Zaid,2018_NR_jeon}. Recent 5G 3GPP specifications follow a revolutionary path of spectrum sharing in licensed and unlicensed spectrum with a significant overhaul of the physical (PHY) layer. 
In this direction, 5G PHY supports features such as tunable sub-carrier spacing, bandwidth and modulation, robust synchronization scheme, and improved channel coding for reliable communication \cite{2017_NR_parkvall,2018_NR_Zaid,2018_NR_jeon,2014_CAHN_hong}. The design of \textit{updated} PHY is important for the upcoming 5G but not sufficient to enable all the above services. It lacks flexibility and significantly depends on upper layers for various decision-making tasks. Real 5G and subsequent 6G i.e. single network enabling heterogeneous services with diverse requirements, cannot be realized without \textit{intelligent} and \textit{reconfigurable} PHY \cite{2018_NR_lien}. Hence, new PHY algorithms, efficient mapping to reconfigurable architectures, and advances in artificial intelligence to embed intelligence need to be explored.

The PHY involves various baseband signal processing tasks such as channel coding, data modulation, resource allocation, layer mapping, precoding, and waveform modulation \cite{2018_NR_lien}. Some of these blocks demand reconfigurable architecture in 5G. For instance, data modulation with different modulation schemes, transmission bandwidth, sub-carrier spacing, interleaving, and precoding parameters can be changed dynamically \cite{2017_NR_parkvall,2018_NR_Zaid,2018_NR_jeon,2018_NR_lien}. Similarly, the orthogonal frequency division multiplexing (OFDM) waveform, which has been successfully deployed in 4G, is selected as a 5G waveform. However, to improve the out-of-band attenuation of OFDM in certain deployment, researchers are exploring various windowing/filtering along with the tunable guard interval \cite{2018_NR_Zaid,2018_NRwaveform_Lie,2015_FOFDM_Zhang,2017_FPGA_kumar,2017_FOFDM_Sasha,2020_FOFDM_Niharika}. Thus, depending on the environment, an on-the-fly switch between PHY parameters is desired. In this direction, making the PHY reconfigurable is important from next-generation network perspective \cite{2020_ROFDM_kumar,2018_sensing_murty,2018_hwsw_drozdenko,2017_ROFDM_pham,2017_hwsw_rihani}.


Reconfigurable PHY can adapt to the changing unknown environment but needs intelligence to learn the environment and dynamically choose the reconfiguration parameters. For example, depending on the channel conditions and throughput desired by underlining services, appropriate PHY parameters such as channel coding rate, modulation scheme, and transmission bandwidth needs to be selected. Furthermore, depending on the selected carrier frequency and given spectrum occupancy, waveform (OFDM, Windowed-OFDM, and Filtered-OFDM) needs to be configured appropriately to meet the desired throughput and interference constraints. With the introduction of a wide range of heterogeneous services, 3GPP has proposed various base-station splits that do not guarantee tight integration of the media access control (MAC) with PHY. Thus PHY with embedded intelligence needs to be explored and is the focus of the proposed work.

In this paper, we design and realize end-to-end intelligent and reconfigurable PHY on Zynq system-on-chip (ZSoC) consisting of an ARM processor as a processing system (PS) and field-programmable gate array (FPGA) as programmable logic (PL) \cite{2013_hwsw_belt}. The hardware-software co-design based profiling approach is used to divide the proposed architecture between PS and PL. The reconfiguration at PS can be easily achieved via software up-gradation, while the dynamic partial reconfiguration (DPR) feature of FPGA is needed for reconfigurability at PL. In the end, we explore a multi-armed bandit (MAB) based online machine learning (OML) algorithms to embed intelligence at the PHY layer \cite{2019_MABbook_Slivkins,2002_MABbook_ozger,2013_TS_agrawal,2019_MABSurvey_Bouneffouf,2012_MABbook_bubeck,2019_darak_jsac}. The main contributions of the paper are summarized as below:

\begin{enumerate}
    \item We design and implement wireless transceiver on Zynq SoC and demonstrate the reconfigurable architecture via DPR. For illustrations, we demonstrate the on-the-fly configuration of the modulation scheme and validate its functionality in various wireless channel conditions along with the advantages over the Velcro approach. In the Velcro approach, multiple blocks are multiplexed in parallel, and all blocks are active even though the output of only one block is used.
    
    \item We design and map the MAB based upper confidence bound (UCB) algorithm and its extensions such as UCB\_V and UCB\_T on ZSoC. The MAB algorithms are based on exploration and exploitation trade-off, and they help o identify optimum parameters (wireless channel, modulation, etc.) without any prior knowledge of the environment.
    
  \item We explore DPR to make the MAB architecture reconfigurable in terms of the number of arms and algorithm type. For instance, depending on the environment, the proposed reconfigurable architecture allows the dynamic switch between MAB algorithms. This is important since a single MAB algorithm may not be suitable in all environments as well as for a given performance and complexity trade-off. 

\item We integrate the MAB architecture with a wireless transceiver and demonstrate intelligent and reconfigurable PHY functionality via the practical wireless application. We show the intelligent on-the-fly selection of modulation scheme and wireless channel to improve throughput.
 
 \item We demonstrate the gain in throughput and bit-error-rate (BER) along with resource utilization and power consumption over conventional approaches. We also discuss the effect of word-length (WL) on the performance and resource utilization of the proposed architecture.
    
\end{enumerate}

The rest of the paper is organized as follows. The review of related works is discussed in Section~\ref{Lit_Review}. Section~\ref{proposed_phy} describes the design details of the intelligent and reconfigurable PHY. Section~\ref{sec:mab} explains the proposed reconfigurable architecture of the MAB algorithm followed by PHY architecture in Section~\ref{sec:phy}. The complete architecture of reconfigurable and intelligent PHY, along with the illustrative demonstration, is presented in Section~\ref{sec:rphy}. Performance analysis and complexity comparison results are discussed in Section~\ref{result_section}. Section~\ref{sec:con} concludes the paper.

\begin{figure*}[!b]
			\vspace{-0.25cm}
			\centering
 			{\includegraphics[scale=0.7]{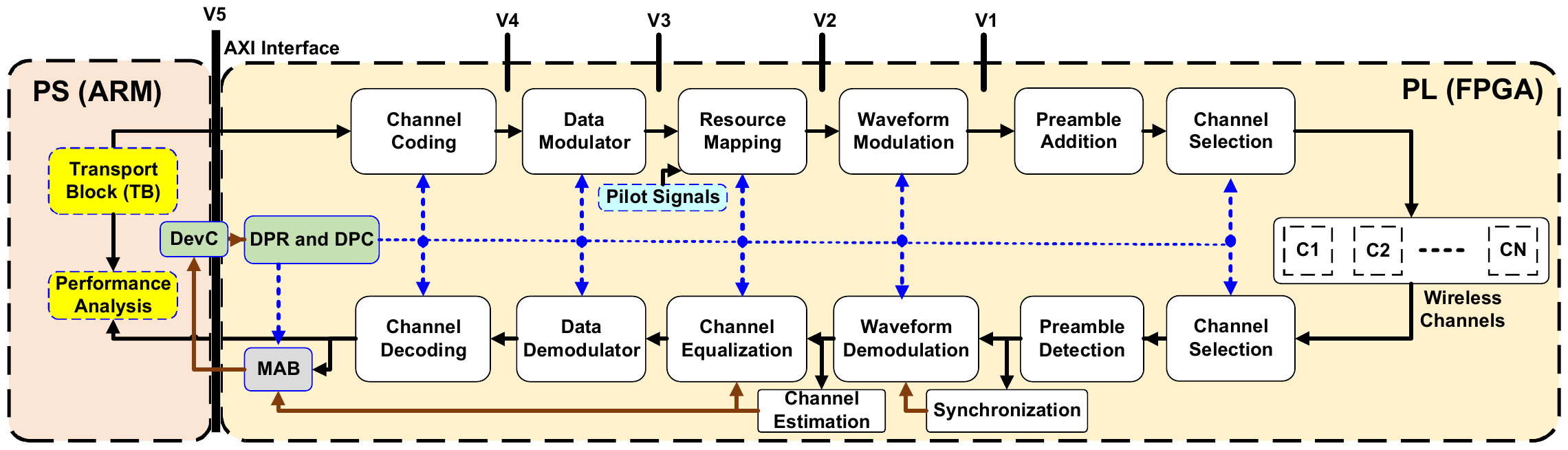}}			\vspace{-0.15cm}
			\caption{Intelligent and Reconfigurable PHY architecture.}
			\vspace{-0.4cm}
			\label{IRPHY}
		\end{figure*}
\vspace{-0.3cm}
\section{Related Work}
\label{Lit_Review}
This section discusses various state-of-the-art works dealing with the design, implementation, and performance analysis of reconfigurable and intelligent PHY.

Wireless PHY has evolved significantly over the years, and there have been significant changes at the algorithm and architecture level. The multi-carrier waveform modulation approaches, such as OFDM and its variants, have been widely deployed in 4G and WiFi networks. Various works dealing with filtering and windowing approaches to improve OFDM performance are discussed in \cite{2020_FOFDM_Niharika,2017_FOFDM_Sasha,2017_FPGA_kumar}. The filter-bank multi-carrier with offset quadrature amplitude modulation (FBMC/OQAM) offers better out-of-band attenuation than OFDM but suffers from high complexity. In \cite{2016_RFBMC_Nadal}, low complexity pipelined architecture for FBMC/OQAM is proposed by optimizing in-built filter architecture resulting in 40-50\% reduction in complexity. Please refer to \cite{2016_NRwaveform_weitkemper} for more details of the various hardware implementation of waveform modulation techniques being considered in 5G.

With the introduction of a wide variety of services and applications, there is a surge of interest in the design of reconfigurable PHY. In the last decade, reconfigurable PHY-based software-defined radios (SDR) were envisioned, which can reconfigure itself on-the-fly, thereby offering flexible and upgrade-able architectures \cite{2016_SDR_Wyglinski,2017_SDRFPGA_Cai,2017_FPGA_kumar}. For the realization of PHY, the first step is to validate its functionality in the real-radio environment, and Universal Software Radio Peripheral (USRP) based testbed is widely used for such performance analysis \cite{2010_USRP_zhang,2018_USRP_zayani,2019_USRP_himani,2020_USRP_bruendl}. These testbeds allow the software-based implementation of the baseband algorithms and RF front-end to validate the performance in a real radio environment. However, it does not consider the effect of quantization due to finite word-length and implementation complexity in terms of area, delay, and power. Thus, various platforms such as ARM processor, FPGA, ASIC, and ZSoC have been explored to implement end-to-end PHY.

The work in \cite{2017_SDRFPGA_Cai} demonstrates the FPGA implementation of PHY via the HDL Coder toolbox of the MATLAB/Simulink and its validation in the real-radio environment. In \cite{2020_ROFDM_kumar}, low complexity and reconfigurable architecture for universal-filtered multicarrier (UFMC) waveform is proposed along with its implementation on FPGA. Detailed performance analysis demonstrates the advantage of FPGA implementation over software (i.e., processor) based implementation. In \cite{2020_sensing_rohit, 2018_sensing_murty}, PHY is integrated with reconfigurable spectrum sensing unit to identify vacant spectrum resources in real-time. Such radios are referred to as cognitive radio (CR). Similar work has also been carried out in \cite{2018_hwsw_shreejith} via low complexity spectrum sensing algorithms. These works are based on mapping algorithms to low complexity efficient architecture on the homogeneous platform but offer limited flexibility and reconfigurability. 

Recently, a hardware-software co-design approach to realize IEEE 802.11a PHY on heterogeneous ZSoC via HDL Coder is presented \cite{2018_hwsw_drozdenko}. Various configurations are explored to identify the appropriate boundary for the division of PHY between PS and PL.  In \cite{2020_FOFDM_Niharika,2017_FOFDM_Sasha}, we developed the end-to-end PHY for the deployment of the air-to-ground communication in $L$-band spectrum and realized it on the ZSoC along with RF front-end. The various filtering and windowing approaches are discussed to improve the transmission bandwidth without compromising the out-of-band interference to $L$-band legacy users. Also, the effectiveness of such approach is demonstrated via in-depth performance analysis for various parameters such as out-of-band attenuation, interference, bit-error-rate, word-length, and complexity in the presence of various RF impairments and air-to-ground specific wireless channels.

As discussed in Section I, efficient design and implementation of PHY are desired, but reconfigurable and intelligent PHY is critical to bring various envisioned services in next-generation networks to reality. Very few works have been done in this direction.  In \cite{2017_hwsw_rihani}, reconfigurable PHY, which allows the dynamic switch between two different communication standards via DPR, is presented, and switching decision is carried out by a vertical handover algorithm based on a scoring system.
In \cite{2015_hwsw_shreejith}, CR with PHY in FPGA and upper layers in the ARM processor is presented, and the digital front-end of the PHY is made reconfigurable via DPR. \cite{2015_hwsw_shreejith} proposed the CR architecture, which does not need separate filtering architectures for channelization and spectrum sensing. The DPR is used to replace a low pass filter for channelizer with the polyphase filter bank for spectrum sensing and vice-versa. Similarly, in \cite{2017_ROFDM_pham}, DPR has been explored to design end-to-end PHY. It can be observed that DPR and hardware-software co-design offer a state-of-the-art approach to make PHY reconfigurable. However, reconfigurable PHY needs intelligence and decision making from the upper layers to decide when and what to reconfigure. In 5G, with the separation of radio unit and distribution unit, PHY and upper layers may not be housed in the same unit, and hence, intelligence at the PHY level is desired. Furthermore, with the recent progress in AI/ML algorithms, hardware-based acceleration of the decision making algorithms of upper layers must meet the area, power, and latency constraints  \cite{2018_CAHN_ozger,2018_CAHN_ali}.

 In this paper, we explore MAB algorithms for the decision making tasks \cite{2019_MABbook_Slivkins,2002_MABbook_ozger,2013_TS_agrawal,2019_MABSurvey_Bouneffouf,2012_MABbook_bubeck,2019_darak_jsac}. MAB algorithms are designed to identify the best arm among several arms in an unknown environment. They guarantee optimal balance between exploration (select all arms a sufficient number of times) and exploitation (select best arm as many times as possible). Popular MAB algorithms include the upper confidence bound (UCB) algorithm and its extensions (UCB\_V and UCB\_T), Kullback-Leibler (KL) divergence based UCB algorithm (KLUCB), and Thompson sampling (TS) \cite{2019_MABbook_Slivkins,2002_MABbook_ozger,2013_TS_agrawal,2019_MABSurvey_Bouneffouf,2012_MABbook_bubeck,2019_darak_jsac}. To the best of our knowledge, none of these algorithms have ever been realized on the SoC. Furthermore, various works have shown that a single MAB algorithm may not guarantee optimal performance under various constraints (performance, area, delay, and power) \cite{2019_MABbook_Slivkins,2002_MABbook_ozger,2013_TS_agrawal,2019_MABSurvey_Bouneffouf,2012_MABbook_bubeck,2019_darak_jsac}. Since the power consumption increases linearly with the number of arms, Velcro approach of parallel implementation of all algorithms for all arms is extremely inefficient. Hence, reconfigurable architecture that allows on-the-fly switching between algorithms and the number of arms is desired, and it is one of the important contributions of the proposed work.

\vspace{-0.3cm}
\section{Intelligent and Reconfigurable PHY}
\label{proposed_phy}
The block diagram of the proposed intelligent and reconfigurable PHY is shown in Fig.~\ref{IRPHY}. The transport block (TB), i.e., the data to be transmitted, is received from upper layers, and it is assumed to be available in memory. The PL, i.e., ARM processor, reads and forwards the TB to the transmitter PHY after receiving the transmit control signal from the upper layer. The TB is processed by channel encoder, data modulator followed by waveform modulation. In the end, the preamble is appended, and the baseband signal is transmitted over one of the selected channels. The received signal on one of the selected channels is processed to detect preamble and identify the OFDM symbol boundary at the receiver. This is followed by channel estimation based on the received and transmitted pilots. Other than channel equalization, the rest of the baseband processing at the receiver is the same as that of the transmitter except that processing is carried out in the reversed direction, and each block performs the inverse operation of the corresponding transmitter block. The performance analysis block compares the transmitted and received TB and calculates the bit-error-rate, latency, and throughput. The DPR and dynamic parameter configuration (DPC) block allow the on-the-fly configuration of various transceivers' blocks. The DPR and DPC blocks are configured by the device configuration (DevC) unit of the PS. The MAB block realizes the learning algorithm and provides input to the DevC block. 

The complete PHY is realized on the ZSoC, and the detailed tutorial explaining the step-by-step process is described in the supplementary \cite{2020_supplementary}. All the blocks in PL are realized using Verilog with wishbone protocol for inter-block communication, while all blocks in PS are realized in C++. PL and PS's communication are established via multiple Advanced eXtensible Interface (AXI) Stream and Lite interfaces. In the next Sections, we discuss the design details of MAB algorithms followed by building blocks of the wireless PHY.

\section{Reconfigurable Architecture for MAB Algorithms}
\label{sec:mab}
MAB algorithms are designed to explore available arms, $K, k\in\{1,2,..K\}$, and exploit the optimum arm in a horizon of size, $N$. The distribution of arms is assumed to be unknown. In our setup, the arms correspond to wireless channels. Since next-generation networks are expected to operate in a licensed, shared, and unlicensed spectrum, complete knowledge of channel availability and fading may not be available and may change over time. Thus, base stations and mobile terminals must-have capability to identify the optimum channel for a given environment for which MAB algorithms offer attractive solution \cite{2019_MABbook_Slivkins,2002_MABbook_ozger,2013_TS_agrawal,2019_MABSurvey_Bouneffouf,2012_MABbook_bubeck,2019_darak_jsac}. In this paper, we focus on efficient mapping of the MAB algorithm on reconfigurable architecture. For illustration, we discuss the UCB algorithm and its extensions. However, the proposed architecture can be extended to other MAB algorithms as long as they are synthesizable on the SoC. 

For wireless communication applications, the time is slotted with a horizon size of $N, n\in\{1,2,.., N\}$ and the horizon is equivalent to the channel coherence time during which its characteristics or distribution remains unchanged. In each time slot, the algorithm can select only one wireless channel denoted by $I_{n}$. The transmitter PHY is then configured to transmit over the selected channel.  At the end of the time slot, the algorithm receives the reward, $R_n$, for the selected channel (i.e., only one feedback in each time slot) from the receiver. Here, the reward is the ratio of received pilot power to the transmitted pilot power. Other types of rewards, depending on channel occupancy and noise variance, can be easily incorporated. 

In the first $K$ time slots of the UCB algorithm, each channel is selected once. After that, in each subsequent time slots, quality factor (QF), $Q(k,n)$ is calculated for each channel. The value of $Q(k,n)$ is given by \cite{2019_MABbook_Slivkins},
		
	\begin{equation}
	\label{qf_ucb}
	    Q_u(k,n) = \frac{X(k,n)}{T(k,n)} + \sqrt{\frac{\alpha \log(n)}{T(k,n)}}
	    \vspace{-0.3cm}
	\end{equation}
	where
	\begin{equation}
	\label{X}
	    X(k,n) = X(k,n-1) + R_{n-1} \cdot \textbf{1}_{\{I_{n-1}==k\}} \quad \forall k
	\end{equation}
\begin{equation}
\label{Rn}
   R_{n-1} = \frac{P_{Rx}}{P_{Tx}} 
\end{equation}
	\begin{equation}
	\label{T}
	    T(k,n) = T(k,n-1) + \textbf{1}_{\{I_{n-1}==k\}} \quad \forall k
	\end{equation}
	
where $P_{Rx}$ is the power of the received pilot, and $P_{Tx}$ is the power of the transmitted pilot in the ${n-1}^{th}$ slot. $\textbf{1}_{cond}$ is an indicator function and it is equal to 1 (or 0) if the condition, $cond$ is TRUE (or FALSE). The parameter, $\alpha$, is an exploration factor that can take any value between 0.5 and 2. Based on calculated QFs in each time slot, the channel with the highest QF is selected, and it is denoted by $I_n$.
	\begin{equation}
	\label{I}
	    I_n = \argmax_k Q(:,n)
	\end{equation}

	Similarly, QF is calculated for UCB\_V and UCB\_T algorithms. We denote $Q_v(k,n)$ and $Q_t(k,n)$ be the QFs in UCB\_V and UCB\_T respectively. Then,
		\begin{equation}
		\label{qf_ucbv}
	    Q_v(k,n) = \frac{X(k,n)}{T(k,n)} + \sqrt{\frac{\alpha_1 \log(n)\cdot V(k,n)}{T(k,n)}} + \frac{\alpha_2 \log(n)}{T(k,n)}
	\end{equation}
	\begin{equation}
		\label{qf_ucbt}
	    Q_t(k,n) = \frac{Y(k,n)}{T(k,n)} - \bigg(\frac{X(k,n)}{T(k,n)}\bigg)^2 + \sqrt{\frac{\alpha \log(n)}{T(k,n)}}
	\end{equation}
		where
		\begin{equation}
		\label{V}
	    V(k,n) = \frac{Y(k,n)}{T(k,n)} - \bigg(\frac{X(k,n)}{T(k,n)}\bigg)^2 \quad \forall k
	\end{equation}
	\begin{equation}
	\label{Y}
	    Y(k,n) = {Y(k,n-1) + (R_n)^2 \cdot \textbf{1}_{\{I_{n}==k\}}} \quad \forall k
	\end{equation}

From a performance and complexity perspective, UCB and UCB\_T select the arm having the highest mean reward (i.e., throughput), and UCB\_T offers lower regret (i.e., higher reward and throughput) than UCB \cite{2019_MABbook_Slivkins,2002_MABbook_ozger}. The UCB\_V algorithm selects the arm based on mean as well as variance. Among the three, UCB\_T is the most complex (area, delay, and power), followed by UCB\_V. The KLUCB offers the best theoretical performance, but it is significantly complex due to the underlining optimization problem \cite{2019_MABbook_Slivkins,2019_MABSurvey_Bouneffouf}. The TS algorithm is difficult to realize on hardware due to an in-built $Beta$ function for which architecture does not exist \cite{2013_TS_agrawal}. In this paper, we focus on UCB, UCB\_V, and UCB\_T algorithms.\footnote{We have also realized KLUCB on ZSoC, but corresponding architectural details are skipped due to space constraints.}  To the best of our knowledge, this work is the first attempt towards the realization and performance analysis of the MAB algorithms on the hardware.
Next, we discuss the mapping of the UCB algorithms to efficient architecture for realization on the SoC. In one-time slot of the MAB algorithm, three operations are performed sequentially: 1) Initialization and parameter update (IPU) to update parameters X (Eq.~\ref{X}), T (Eq.~\ref{T}), Y (Eq.~\ref{Y}) and $n$, 2) QF calculation for each channel based on the updated parameters (Eq~\ref{qf_ucb}, Eq~\ref{qf_ucbv} and Eq~\ref{qf_ucbt}), and 3) Channel selection using updated QFs (Eq.~\ref{I}).

 \vspace{-0.3cm}
 \subsection{Initialization and Parameter Update (IPU) Block}
\begin{figure}[!h]
 \vspace{-0.3cm}
 \includegraphics[width=0.9\linewidth]{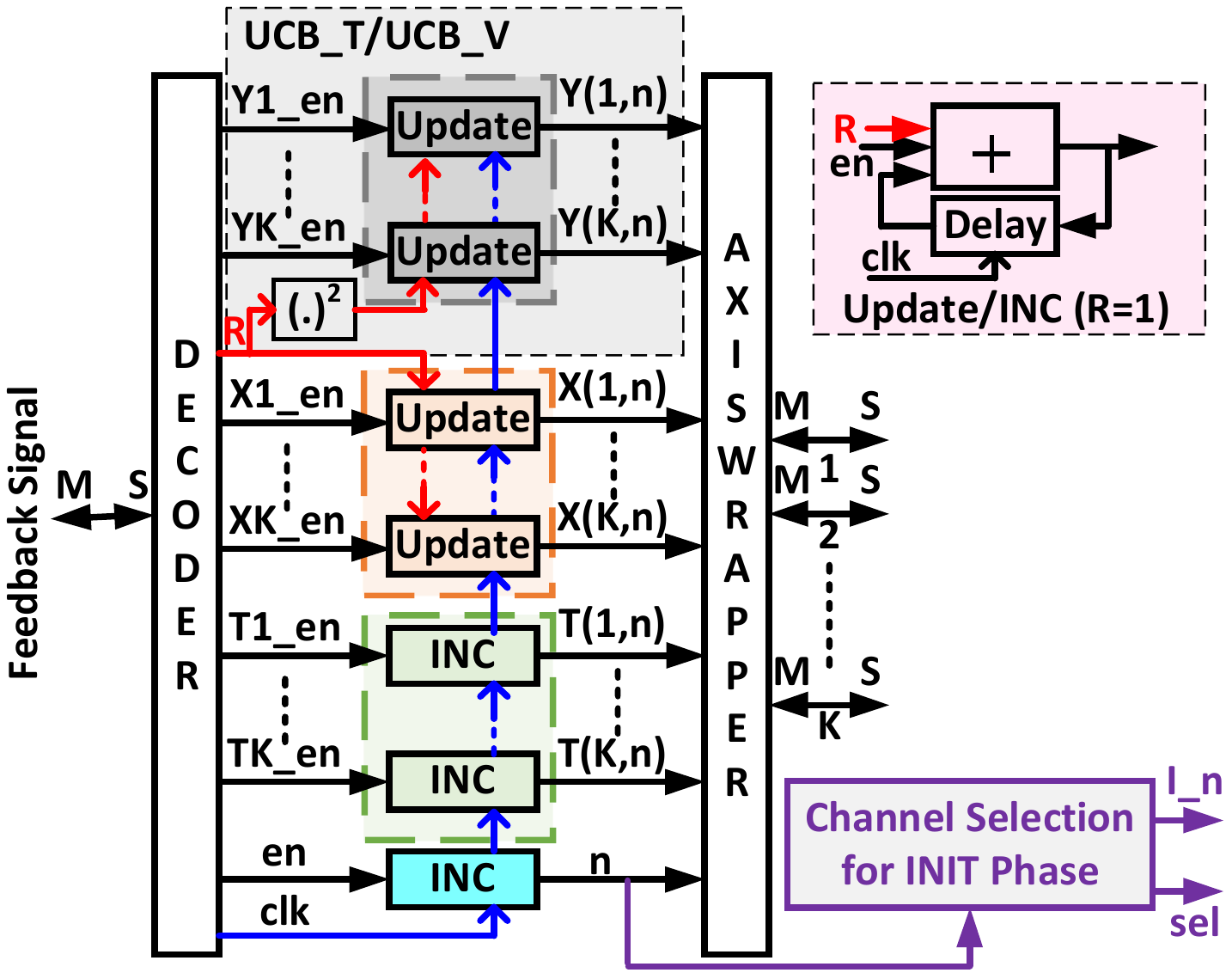}
 \vspace{-0.2cm}
 \caption{Initialization and Parameter Update block}
 \vspace{-0.3cm}
 \label{pua}
\end{figure}

  The IPU block operates either in the initialization (INIT) mode or learning (LEARN) mode. The INIT mode is of duration $K$ slots in which all parameters $\{X, T,n, Y\}$ are set to zero in the first slot and updated in the subsequent slots as per Eq.~\ref{X}, Eq.~\ref{T}, and Eq.~\ref{Y}. In the INIT mode, each channel is selected once via pseudo-random sequence generator of length $K$, and hence, QF calculation and channel arm selector blocks are not enabled. After the first $K$ slots, the algorithm enters into the LEARN mode, where the IPU block updates the parameters based on the information received via feedback signal. The IPU block architecture, shown in Fig.~\ref{pua}, consists of adders, delays, decoders, and a pseudo-random sequence generator.
  
\begin{figure}[!h]
 	\vspace{-0.3cm}
 	\includegraphics[width=0.45\textwidth]{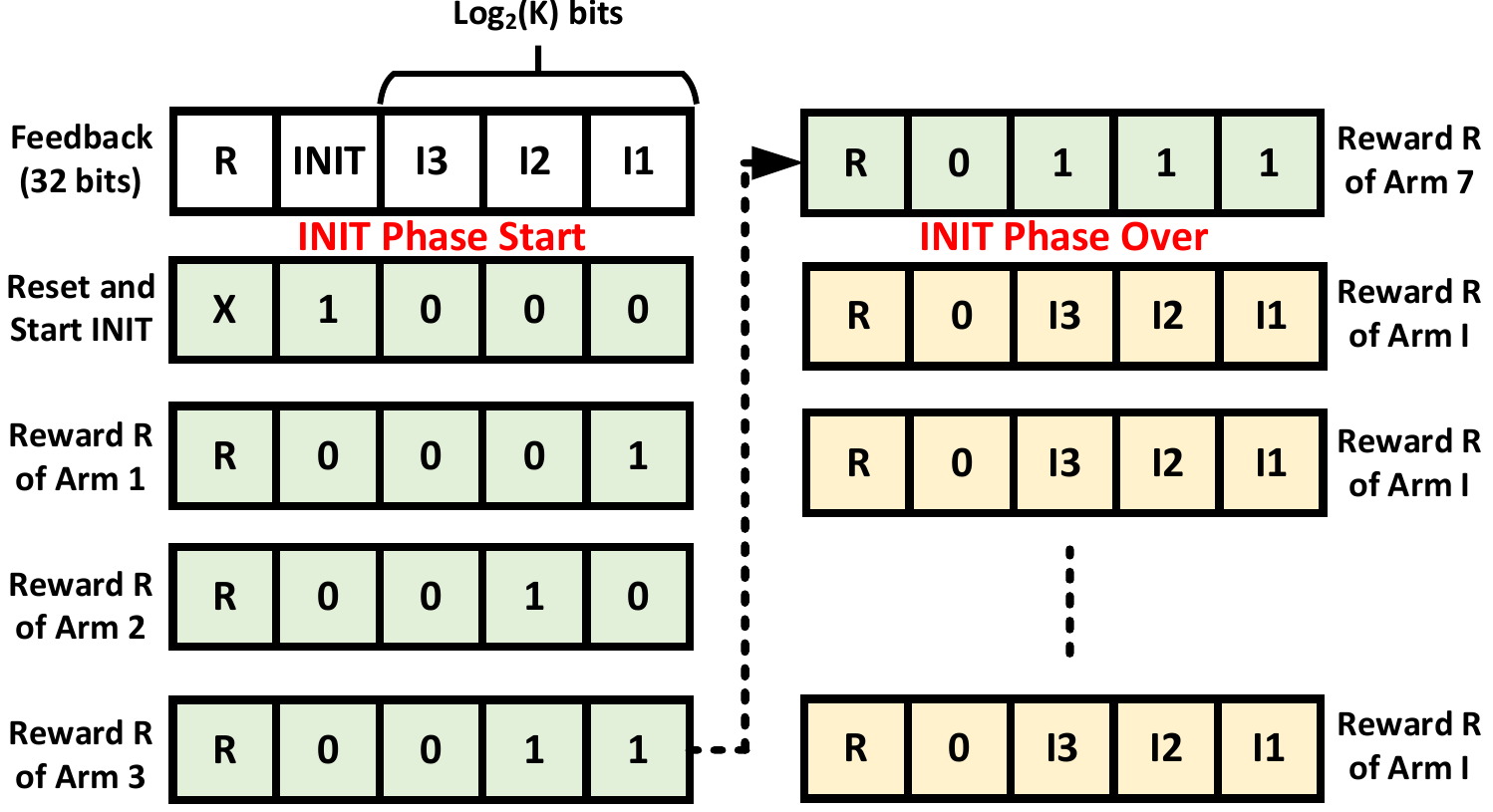}
 	\vspace{-0.1cm}
 	\caption{ Feedback signal format which is an input to initialization and parameter update block.}
 	\vspace{-0.3cm}
 	\label{feedback}
 \end{figure}

 The feedback signal is the input to the IPU block, and it is received from the wireless PHY. It gives the information about the status of the channel selected in the previous time slot, $I_{n-1}$, mode (INIT or LEARN), and the corresponding reward, $R_{n-1}$. For $K$ channels, we need $\log_2(K)$ bits, and since AXI registers are of 32 bits, the rest of $(32-\log_2(K)-1)$ bits represent the reward (Eq.~\ref{Rn}) which can take any value between 0 and 1. Higher the reward, the better the quality of the channel. The feedback signal format with illustrative INIT mode operation for $K=3$ is shown in Fig.~\ref{feedback}. For easier understanding, the channels are shown to be selected in the deterministic order. However, the channels are selected in a pseudo-random manner using a pseudo-random sequence generator such that each channel is selected once in the INIT mode.
  
  The feedback signal is decoded, and corresponding enable signals are generated, as shown in Fig.~\ref{pua}. For instance, \textit{en} is generated at the beginning of every slot.  Based on the channel selected in the previous time slot, appropriate $T_{k\_en}$, $Y_{k\_en}$ and $X_{k\_en}$ are generated. For example, if the second channel is selected, then $T_{2\_en}$, $Y_{2\_en}$ and $X_{2\_en}$ are HIGH and rest are LOW. Also, $R$ is the reward received in the previous slot, which is added to $X_{k\_en}$ in the update block, as shown in Fig.~\ref{pua}. Similarly, the squared value of $R$ is used for updating $Y_{k\_en}$. Since $T_{k\_en}$ and $n$ are incremented by 1, INC block is used which is same as update block with $R=1$. Note that the channel selection block is enabled only when $n\leq K$.

 \vspace{-0.3cm}
\subsection{Quality Factor Calculation}
\begin{figure}[!h]
		\vspace{-0.25cm}
		\centering
		\includegraphics[width=0.45\textwidth]{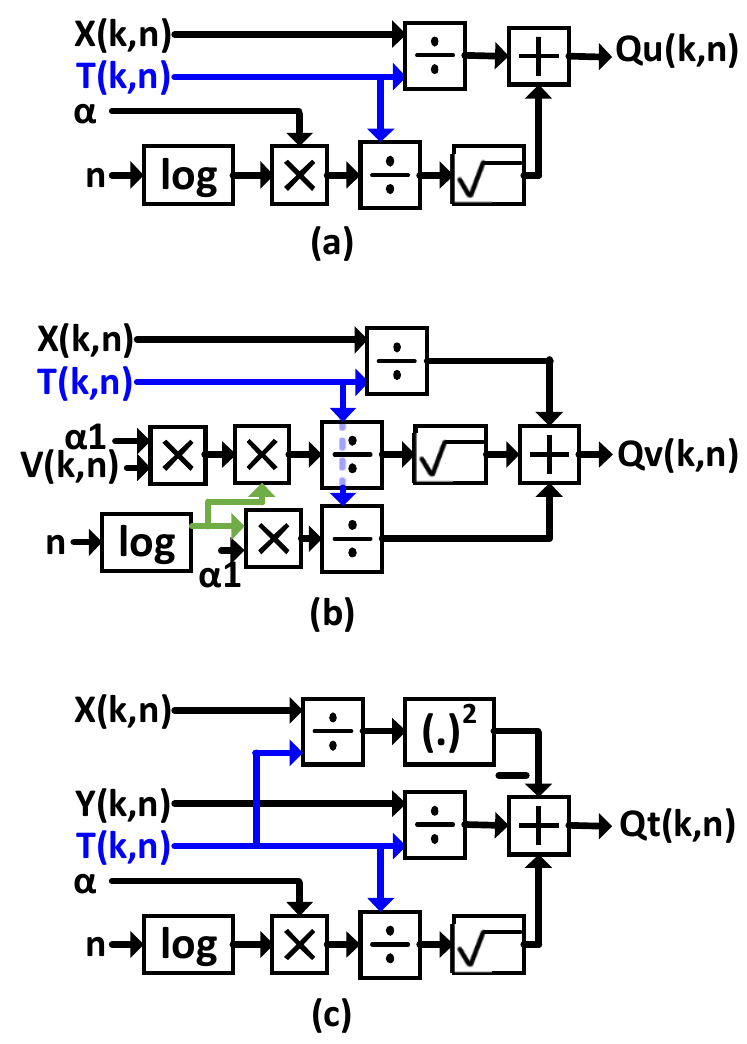}
		\vspace{-0.25cm}
		\caption{ QF calculation block for (a) UCB algorithm, (b) UCB\_V algorithm, and (c) UCB\_T algorithm. Note that QF calculation is done using floating as well as fixed-point arithmetic and AXI handshake signals are omitted for maintaining clarity of figures.}
		\label{ucb}
		\vspace{-0.2cm}
	\end{figure}

The next step after IPU is to calculate the QF for every channel using the parameters updated by IPU. The architectures QF calculation of single channels corresponding to UCB, UCB\_V, and UCB\_T algorithms are shown in Fig.~\ref{ucb}. The architectures are based on Eq.~\ref{qf_ucb}, Eq.~\ref{qf_ucbv} and Eq.~\ref{qf_ucbt}. Note that we have omitted AXI handshake signals to maintain the clarity of the architecture. Also, QF of all channels is done in parallel in the PL, while PS implementation results in sequential implementation leading to high latency, as discussed later in Section~\ref{result_section}. 

All inputs in Fig.~\ref{ucb} are received via the AXI4-stream interface with the IPU block. Similarly, there is a single AXI4-stream interface at the output, which is interfaced with the channel selection block. The IPU and QF calculation block can be combined into one block. However, since the feedback signal is received from PS via the AXI4-Lite interface, resultant latency is significantly high due to four handshakes compared to a single handshake in the current architecture. All the architectures in Fig.~\ref{ucb} are realized using floating as well as fixed-point arithmetic. We have explored various combinations of WL, and corresponding results are discussed later in Section~\ref{result_section}.

	\vspace{-0.3cm}
\subsection{Channel Selection}
In each slot, the channel with maximum QF value is selected by comparing the QF values received from the QF calculator block, and the corresponding architecture for $K=4$ is shown in Eq.~\ref{I}. The output of the selection block is channel index $I_n$, which is used by wireless PHY for the data transmission in the rest of the time slot.

\begin{figure}[!h]
 \vspace{-0.3cm}
 \includegraphics[width=0.9\linewidth]{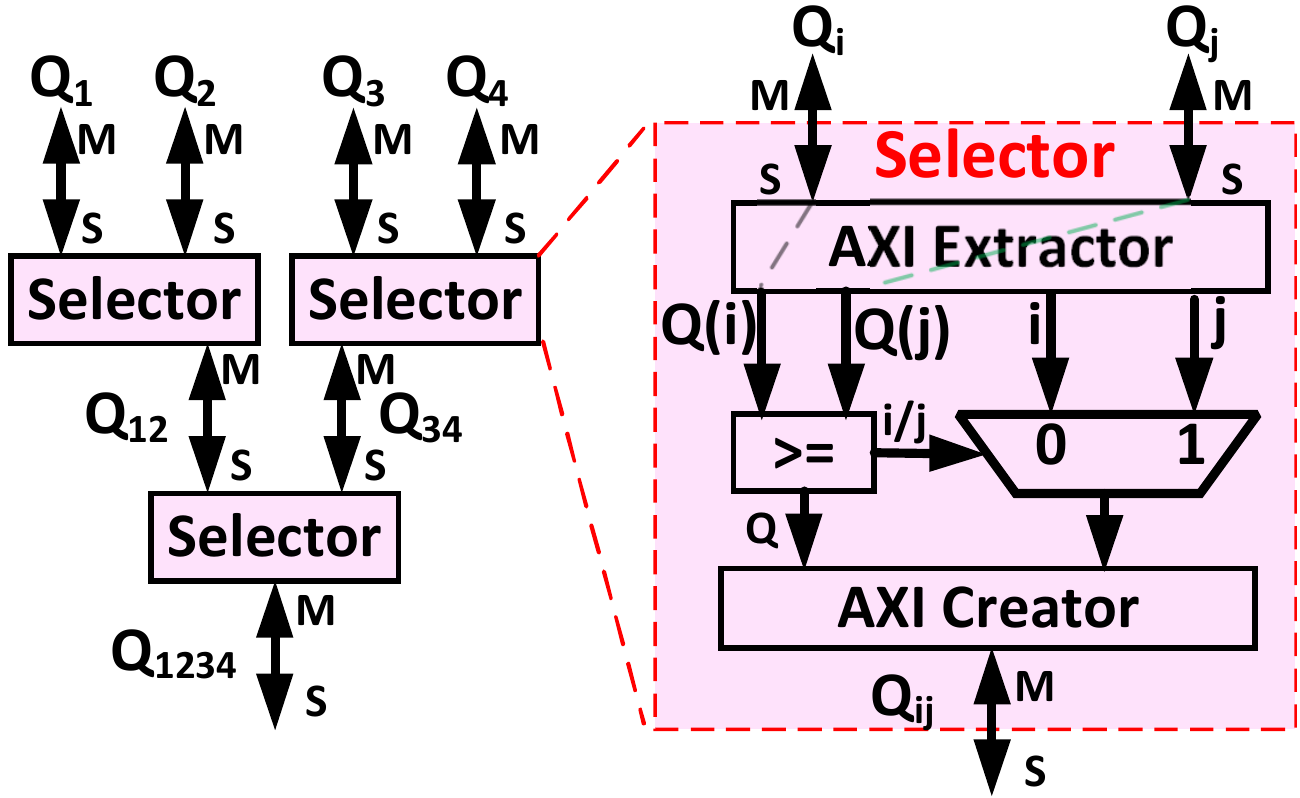}
 \vspace{-0.1cm}
 \caption{QF based channel selection block.}
 \label{sel}
 \vspace{-0.3cm}
\end{figure}

\vspace{-0.3cm}
\section{Wireless PHY Architecture}
\label{sec:phy}
This section briefly discusses the design details of various transmitter and receiver PHY blocks on PL via Verilog-based IPs. We begin with the wishbone protocol details, which is used for communication between multiple blocks inside PL.
\vspace{-0.3cm}
\subsection{Wishbone Protocol}
Wishbone is an industry-standard communication protocol enabling a common interface for all the system blocks irrespective of their internal functionality. Similar to other interfaces such as AXI, the wishbone protocol allows communication between master and slave blocks. For a given interface, one block is master, and another block is a slave. The master must initiate all transactions (read, write, and block transfer). For example, in Fig.~\ref{IRPHY}, for the interface between channel coding and data modulator blocks, the former is the master, and the latter is the slave. On the other hand, the data demodulator is master, and channel decoding is a slave for the interface between the two in the receiver. 
The important signals of the wishbone interface are given below \cite{2017_ROFDM_pham}:

\begin{itemize}
\subsubsection{Common Signals}
\item CLK\_O:- This is a clock input signal generated using the clock output of the PS (i.e. ARM processor). In our architecture, the clock given to each block may not be the same and need to be updated depending on the given configuration. For example, the clock frequency of 16-QAM is twice that of QPSK. 
\item RST\_O:- This is reset input signal and it is connected to the reset signal output of the PS (i.e. ARM processor). This is used to bring the transmitter and receiver PHY to the known state after every time slot.

\item DAT\_I (Master Input) and DAT\_O (Slave output):- Both enables the data transfer from slave to master. Similarly, DAT\_O (Master Output) and DAT\_I (Slave input) enable the data transfer from master to slave.

\item ACK\_I (Master input) and ACK\_O (Slave output):- They form an acknowledge signal which is asserted by a slave to indicate the termination of a normal bus cycle.

\item STB\_O (Master output) and STB\_I (Slave input):- When a master is ready to transfer the data, it asserts the STB\_O signal. Slave responds to the received strobe signal via acknowledgment signal. 

\item WE\_O (Master output) and WE\_I (Slave input):- It is asserted during the write cycle and de-asserted during the read cycle.
\item CYC\_O (Master output) and CYC\_I (Slave input):- It indicates the valid bus cycle and remains asserted for all data transfers in a given bus cycle. 
\end{itemize}
\vspace{-0.3cm}
\subsection{Transmitter PHY}
As shown in Fig.~\ref{IRPHY}, the transmitter PHY consists of various baseband signal processing blocks to convert the upper layer TB into OFDM samples for transmission over the channel. In the transmitter PHY, MAB, and DevC blocks decide the channel to be selected in a subsequent time slot, and based on the learned distribution of the selected channel, one of the modulation schemes (16-QAM or QPSK) is selected. Thus, input to the transmitter PHY is the modulation scheme, channel, and TB of appropriate size. For example, since 16-QAM maps four bits to one symbol and QPSK maps two bits to one symbol, the size of TB is twice for 16-QAM than that of QPSK. 

\begin{figure}[!h]
 \centering
 \includegraphics[width=\linewidth]{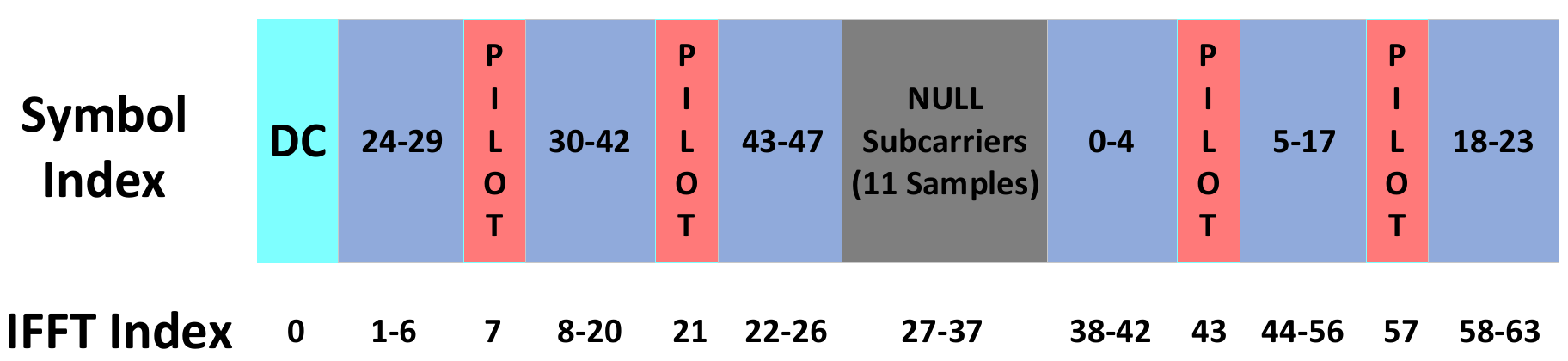}
 \caption{Symbols to resource element mapping for the proposed architecture.}
 \label{re}
 \vspace{-0.1cm}
\end{figure}

In the proposed work, the transmitter PHY is based on IEEE 802.11a specifications with 64 sub-carriers (can be extended to any other size), out of which 48 are data sub-carriers, 4 pilot sub-carriers, and 12 are null sub-carriers. The position of these sub-carries is shown in Fig.~\ref{re}. For every OFDM symbol, the channel coding block generates 98 bits for QPSK or 196 bits for 16-QAM modulation schemes. These bits are converted to the symbols by the data modulation block. The proposed data modulation follows 3GPP specifications for 5G PHY and generates a complex symbol with Inphase (I) and Quadrature (Q) components. The Modulation mapping of input of single-channel (I and Q) in a signed 16-bit Q1.15 format is shown in Table~\ref{mod_map}.

\begin{table}[!h]
\centering

\caption{Mapping of input bits using Q1.15 format}
\renewcommand{\arraystretch}{1.3}
\resizebox{\linewidth}{!}{

\begin{tabular}{|c| c| c| c|} 
 \hline
 Modulation Scheme & Bit & Mapped value in Hex & Signed Mapped Value\\ [0.5ex] 
 \hline\hline
 QPSK & 0 & A57E & -0.7071 \\ 
 QPSK & 1 & 5A82 &  0.7071 \\
 16-QAM & 00 & 8692  & -0.9485 \\
 16-QAM & 01 & 287A & 0.3162 \\
 16-QAM & 10 & D786 & -0.3162 \\
 16-QAM & 11 & 796E &  0.9485\\ [1ex] 
 \hline
\end{tabular}
}
\label{mod_map}
\vspace{-0.3cm}
\end{table} 

Similarly, pilot symbols are BPSK modulated. The phase of the pilot signal depends on its sub-carrier index and is fixed. Thus, pilot signals are pre-generated and stored in block memory of PL. As per Q1.15 format, the value of the pilot symbol is 0x7FFF and 0x8001 for +1 and -1, respectively. The modulated data and pilot symbols are mapped to appropriate sub-carriers by the resource mapping block, which acts as a serial-to-parallel converter.  This is followed by a 64-point inverse Fast Fourier transform (IFFT) using Xilinx optimized IFFT block. The IFFT output is serialized to obtain 64 time-domain samples, which are then extended to 80 samples via cyclic prefix (CP) addition. Here, the last 16 samples are appended initially, which in turn reduces the impact of inter-symbol interference due to multi-path channels. 

Since the transmitter and receiver PHY layers are not synchronized, receiver PHY needs to identify symbol and frame boundary, estimate and compensate the frequency offset, and perform channel estimation and equalization. In cellular communication such as 5G, synchronization and reference signals are used, while in WiFi IEEE 802.11 standard, preambles (long and short) are used. In this work, we have implemented the preamble based approach. As per IEEE 802.11 standard, a short preamble consists of 10 slots, each composed of 16 samples, while a long preamble consists of two slots of 64 samples each along with CP of 32 samples. Thus, the total preamble occupies 320 samples followed by multiple data payload, each consisting of 80 samples. In our implementation, preamble samples are fixed and are stored in block memory of PL. The preamble addition block includes the scheduler, which maps the appropriate preamble in the desired time slots.

\vspace{-0.3cm}
\subsection{Wireless Channel}
The transmit PHY selects and transmits on one of the $N$ channels, as shown in Fig.~\ref{IRPHY}. Each channel is modeled as a multi-path fading channel with complex symmetric Gaussian distribution. The mean and variance of $n^{th}$ channel is denoted as $\mu_n$ and $\sigma_n^2$, respectively. In each time slot, the MAB algorithm selects the channel, and the sample (fading coefficient) is generated based on the distribution of the selected channel. In the proposed architecture, we have deployed the AXI-Lite interface between PS and PL to convey the selected channel's index along with the fading coefficient to the PHY. In the PL, complex multiplication is performed between the transmit samples and fading coefficient, which is then passed to the receiver PHY.








\begin{figure*}[!b]
\centering
\vspace{-0.3cm}
\captionsetup{justification=centering}
\includegraphics[scale=1.2]{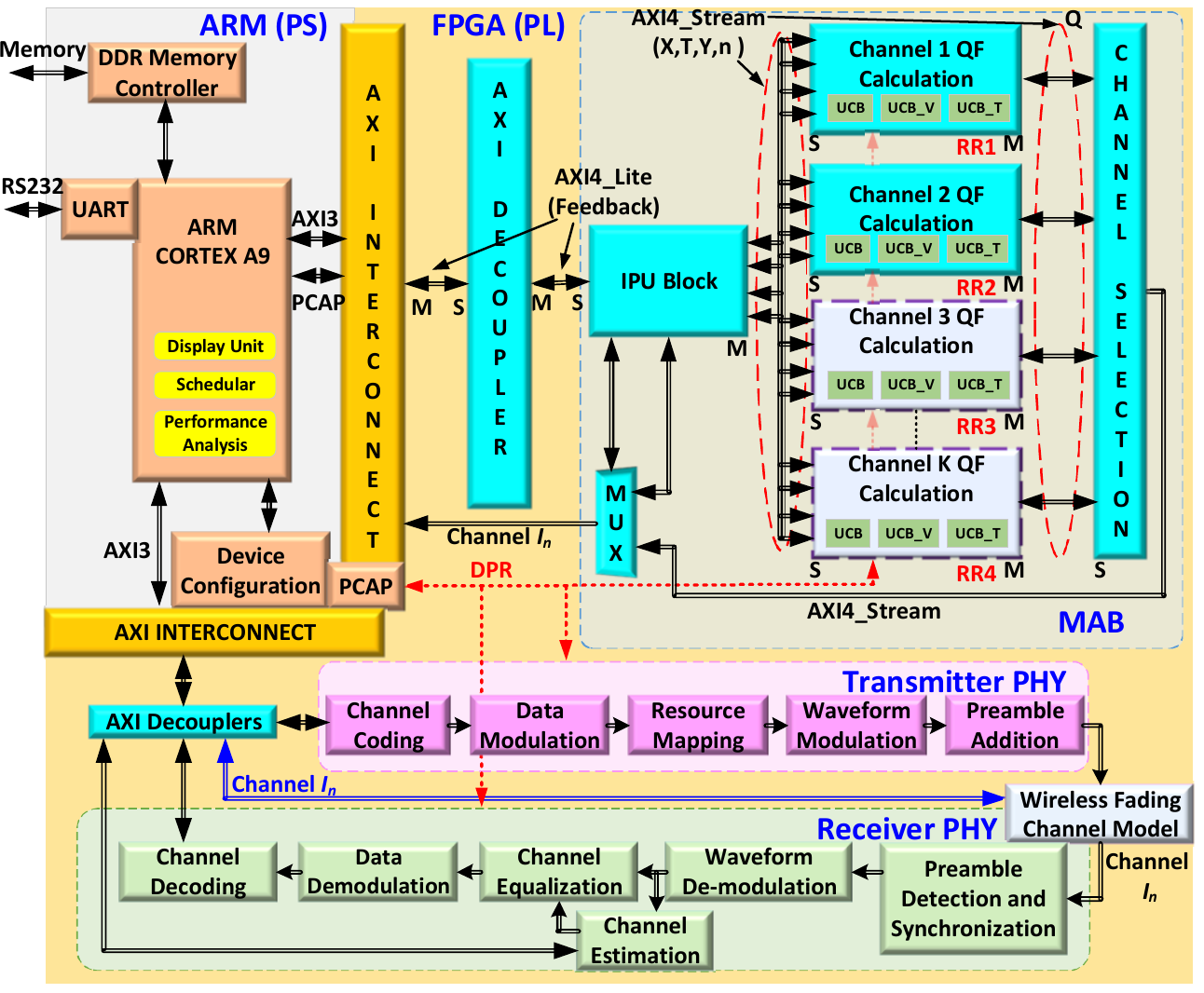}
\vspace{-0.2cm}
\caption{Proposed intelligent and reconfigurable PHY architecture mapped on the ZSoC via hardware software co-design.}
\label{finalArch}
\end{figure*}
\vspace{-0.3cm}
\subsection{Receiver PHY}
The received samples are processed in the synchronization block, followed by waveform demodulation, channel estimation, channel equalization, and data demodulation. 

The first task of the receiver is to detect the preamble and used the same for synchronization tasks. For example, a short preamble is used for the start of the frame detection, and coarse frequency offset estimation, while a long preamble is used for fine frequency offset estimation. It is also used later for channel estimation. 

For frame detection, an auto-correlator is designed with a window length of 32 samples, and auto-correlation is performed between a block of samples separated by 16 samples. This is also known as CP detection. The output of the auto-correlation block is given by\cite{sync,sync1,1}:
\begin{equation}
\label{P}
P[i]=\sum_{m=0}^{31} (r^{*}[i+m] r[i+m+16]),
\end{equation}
where * indicates complex conjugate operation and $i$ is the time index. The output of the auto-correlator is normalized as given below \cite{sync,sync1}:
\begin{equation}
M[i]=\frac{|P[i]^2}{(R[i]^2)}.
\label{time matric}
\end{equation}
where
\begin{equation}
\label{R}
R[i]=\sum_{m=0}^{31}|r[i+m+16] |^2,
\end{equation}
Then, vector $M$ is passed through the comparator to generate the binary signal, which goes High when the corresponding value of $M$ is above the threshold (0.75 in our case).  Ideally, we get one low-to-high transition in a single frame for long preamble and one low-to-high transition for every OFDM symbol carrying data. These transitions are used to identify the start of the OFDM symbol, and hence, CP can be easily removed. Before the waveform demodulation, short preamble based auto-correlation is used to estimate coarse frequency offset, and long preamble based auto-correlation is used to estimate a fine frequency offset \cite{sync,sync1}. After offset corrections, the OFDM symbol corresponding to the long preamble and data are passed through 64-point FFT. The FFT output of the preamble is used for channel estimation, which is nothing but 64 complex multiplication operations between transmitted and received preamble symbols. Data demodulation block extracts the bits from the FFT outputs via hard decoding. In the end, TB is formed, which is passed to the PS for the performance analysis.



\vspace{-0.3cm}
\section{Intelligent and Reconfigurable PHY on ZSoC}
\label{sec:rphy}
In this section, we present the complete architecture of the proposed intelligent and reconfigurable PHY realized on the ZSoC consisting of PS (Dual-core ARM processor) and PL (7-series FPGA). The architecture, shown in Fig.~\ref{finalArch}, mainly consists of MAB algorithm, transmitter, and receiver blocks in PL along with scheduler and performance analysis in PS. In \cite{2020_supplementary}, the detailed tutorial to build the architecture in Fig.~\ref{finalArch} followed by enabling the DPR to make it reconfigurable along with source code is given. 

As discussed in Section~\ref{sec:mab}, the MAB algorithm consists of three sub-blocks, as shown in Fig.~\ref{finalArch}. At the beginning of each experiment, the PS resets the algorithm via feedback signal over the AXI-Lite interface ($X=0, Y=0, T=0,n=1$). In the subsequent $K$ time slots, the IPU bypasses the QF calculation and channel selection blocks to execute the INIT mode and finds the index of the selected channel, $I_n$. At the beginning of each time slot, the PS generates the feedback signal containing the reward information received from the PHY. This information is used by IPU to update the MAB parameters. After the INIT mode, QF and channel selection blocks are enabled, which select the appropriate channel in the LEARN mode. We have explored the complete or partial realization of MAB in the PL. Please refer to Section~\ref{result_section} for the performance analysis of these architectures.
	
For the architecture in Fig.~\ref{finalArch} with $K_{max}=4$, four reconfiguration regions (RR), i.e., the region whose functionality can be changed on-the-fly via DPR, are shown. Since each region can be configured with blank, UCB, UCB\_V or UCB\_T QF blocks, corresponding partial bit-streams can be stored either in on-board DDR memory or SD card. Via bare-metal application deployed on the ARM processor, the desired bit-streams are sent to the FPGA for appropriate RR configuration using the device configuration (DevC) unit of the PS. The FPGA's DPR property achieves this, and we specifically employed ARM processor-controlled DPR via processor configuration access port (PCAP).
	
As discussed previously, the transmitter consists of various signal processing blocks that map the TB to OFDM symbols and transmit them over one of the $K$ channels. The transmitter architecture is reconfigurable because it dynamically adapts the modulation scheme based on the selected channel. This is also achieved via the DPR property of the FPGA configured through the ARM processor. The transmitter maps the output on the channel chosen by the MAB algorithm.

The receiver receives the transmitted OFDM samples on the chosen channel. As discussed before, the receiver estimates the channel quality based on the transmit and receive pilot powers and forwards it to the MAB algorithm via PS. The receiver decodes the received symbols and forms the TB, which is then compared with transmitted TB to obtain BER and throughput in the performance analysis unit of the PS. 

Next, we demonstrate the proposed architecture's functionality for the adaptive modulation scheme and channel selection application. For illustration, we consider two modulation schemes: 1) QPSK and, 2) 16-QAM and $K=5$ channels. In QPSK, one symbol comprises two bits, while in 16-QAM, one symbol contains 4 bits. Thus, when a channel is good (i.e., the noise power is less), 16-QAM is preferred to maximize the throughput. However, at high noise power, QPSK is preferred to achieve moderate throughput with good reliability. Since channels are unknown, PHY needs to learn them accurately before choosing the modulation scheme. We develop the UART based interface to display the results of each experiment.

 \begin{figure}[!h]
\centering
\includegraphics[width=0.45\textwidth]{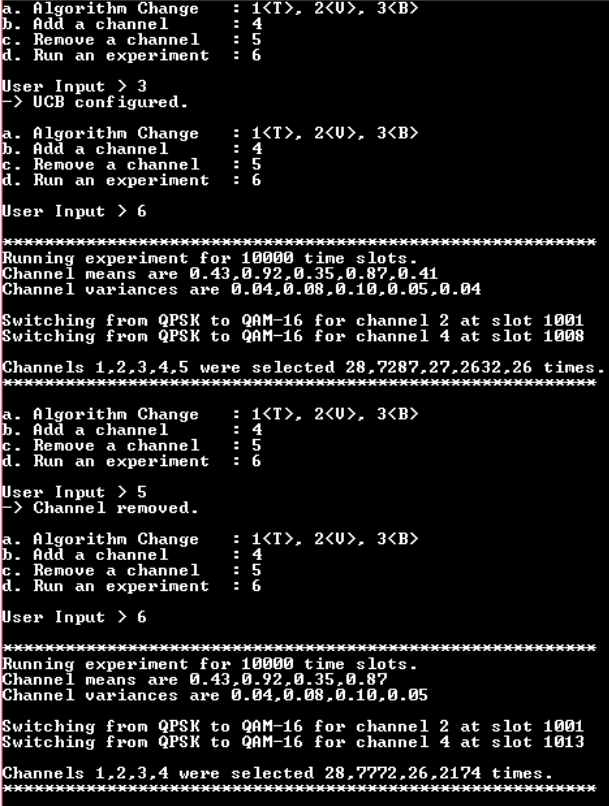}
\vspace{-0.1cm}
\caption{UART based snapshot of baremetal application on ZSoC demonstrating the functionality of proposed PHY.}
\label{snap1}
\vspace{-0.3cm}
\end{figure}

 In Fig.~\ref{snap1}, we consider $K=5$ channels with average throughput of $\mu=\{0.43, 0.92, 0.35, 0.87, 0.41\}$ and variance of $\sigma^2=\{0.04, 0.08, 0.1, 0.05, 0.04\}$. It is assumed that PHY does not have any prior knowledge of the channel. Thus, the MAB algorithm aims to identify the optimum channel via exploration-exploitation trade-off and choose an appropriate modulation scheme to maximize the throughput. Fig.~\ref{snap1} demonstrates the reconfigurable feature of the architecture where DPR based on-the-fly switch between UCB, UCB\_V, and UCB\_T algorithms is shown. Similarly, the architecture can be reconfigured for any $K\leq 6$. For each experiment, the horizon is of size 10000-time slots, and it can be seen that the algorithm selects the channel with higher average throughput more times, i.e., channel 2 is selected in 7287-time slots out of 10000-time slots. Fig.~\ref{snap1} also demonstrates an on-the-fly switch between the number of channels where the last channel is made unavailable, and in the new experiment, channel 2 has been selected 7772 times.

\begin{figure}[!h]
\centering
\includegraphics[width=0.45\textwidth]{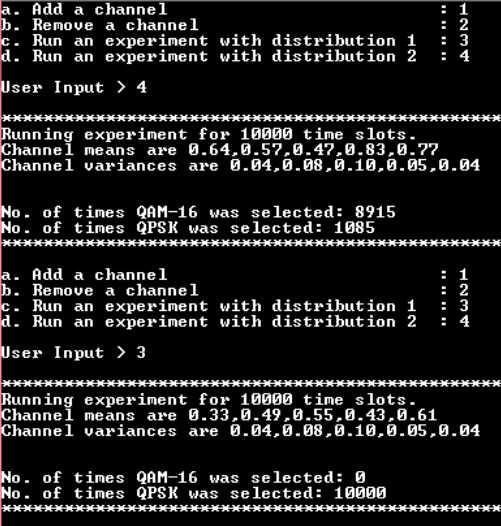}
\vspace{-0.1cm}
\caption{UART based snapshot of baremetal application on ZSoC demonstrating the functionality of proposed PHY.}
\label{snap2}
\vspace{-0.3cm}
\end{figure}

 Similarly, when channel statistics are changed, the architecture intelligently chooses the appropriate configuration (modulation scheme and channel) without any prior information. Fig.~\ref{snap2}, demonstrates that with the change in channel distribution, intelligent on-the-fly adaptation to the desired modulation scheme is accomplished without any manual intervention. For distribution 1 with good channels, 16-QAM is preferred, while for average channel conditions, QPSK is preferred due to the high BER of the 16-QAM. Please refer to Section~\ref{result_section} for details.

\begin{figure}[!h]
\centering
\vspace{-0.1cm}
\includegraphics[width=0.45\textwidth]{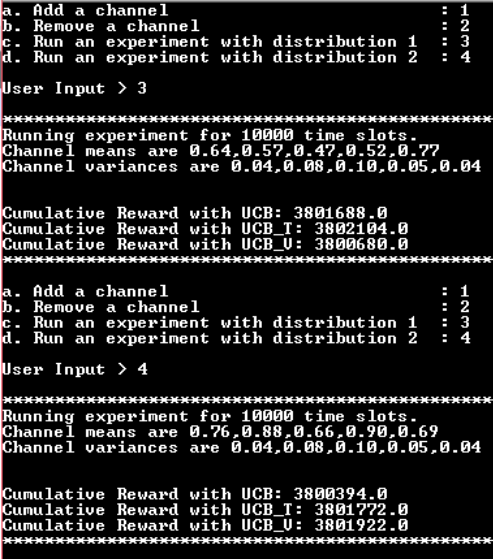}
\vspace{-0.1cm}
\caption{UART based snapshot of baremetal application on ZSoC comparing MAB algorithms.}
\label{snap3}
\vspace{-0.3cm}
\end{figure}

Next, we demonstrate the need for reconfigurable architecture, which can switch between MAB algorithms. For illustration, the total reward at the end of the horizon size of 10000-time slots is compared. Here, the reward of the particular slot is assumed to be equal to the total number of correctly received bits in that time slot. As shown in Fig.~\ref{snap3}, UCB\_T and UCB\_V algorithms offer higher rewards in the first and second cases, respectively. The proposed reconfigurable architecture does not need parallel implementation of all MAB algorithms, thereby significantly reducing complexity. In the current architecture, the user needs to select the algorithm at the beginning of the horizon since there is no existing approach to decide the appropriate MAB algorithm for a given unknown environment. 

\vspace{-0.3cm}
\section{Performance and Complexity Analysis}
\label{result_section}
In this section, we validate the functional correctness of the proposed architecture and the complexity comparison. We begin with the performance analysis of UCB architecture.

\vspace{-0.3cm}
\subsection{Performance Analysis of UCB algorithms}
In this section, we validate the functional correctness of the proposed UCB architecture for different WL. Since the purpose of the UCB algorithms is to assist in the selection of the optimal channel via exploration-exploitation trade-off, we study the channel selected by UCB algorithm in various channel environment. We first consider $K$ = 5, $N$ = 10000 and two different channel distributions: \\
\noindent 1) \{$\mu_1, \sigma_1^2$\} = \{(0.5,0.01), (0.8, 0.02), (0.61, 0.08), (0.45, 0.06), (0.9, 0.07)\} \\
\noindent 2) \{$\mu_2, \sigma_2^2$\} = \{(0.55,0.04), (0.48, 0.14), (0.8, 0.2), (0.72, 0.3), (0.61, 0.2)\}. 

\begin{figure}[!h]
\centering
\subfloat[]{\includegraphics[width=0.475\textwidth]{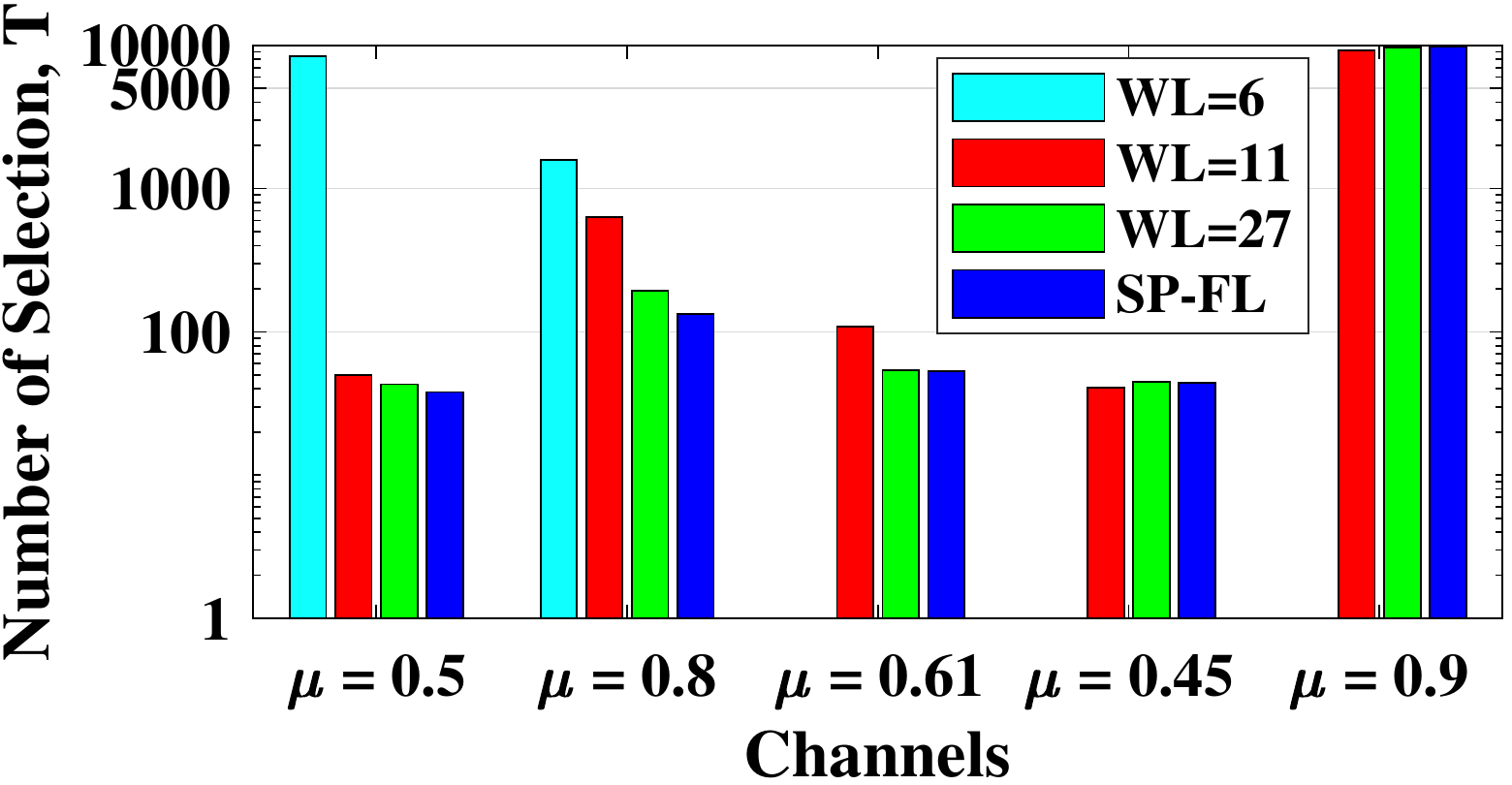}%
 \label{MABK51}}\\
 \vspace{-0.3cm}
\subfloat[]{\includegraphics[width=0.475\textwidth]{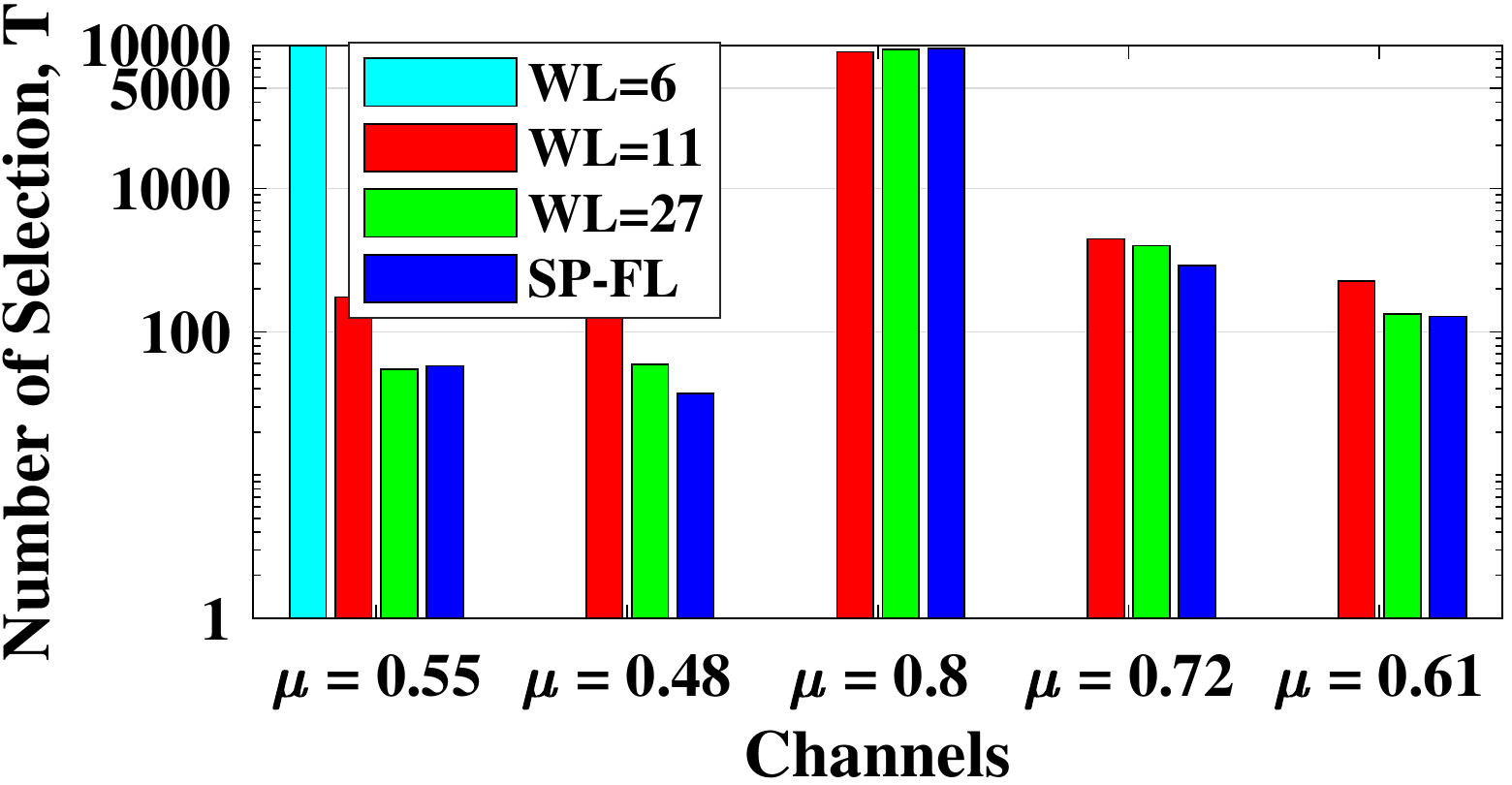}%
 \label{MABK52}}
\caption{Channel selection by UCB algorithms after $N=10000$ time slots with $K$ = 5, (a) $\mu_1$, and (b) $\mu_2$ }
\label{MABK5}
\end{figure}

Here, the channel is assumed to have Gaussian distribution with a mean $\mu$, which corresponds to the channel's average gain, and $\sigma^2$ is the variance. In Fig.~\ref{MABK51} and Fig.~\ref{MABK52}, we consider three different fixed-point WLs along with a single-precision floating-point (SP-FL) architectures. It can be observed that the proposed SP-FL architecture consistently selects the fifth and third channels, which are the optimum channel (i.e., a channel with the highest gain) in distribution $\mu_1$ and $\mu_2$, respectively. Note that all results are obtained after averaging over ten different experiments to consider the non-deterministic nature of the online machine learning algorithms. The values on the $y$-axis are shown on the logarithmic-scale for better clarity. Thus, it can be observed that the proposed UCB architecture identifies the optimum channel quickly, i.e., it optimizes exploration-exploitation trade-off by minimizing the selection of non-optimum channels. Among the fixed-point architectures, the performance of the architectures with WL of 27 and 11 is acceptable as both select the optimum channels a higher number of times. However, for WL of 6, the UCB algorithm's functionality is not correct as it fails to identify the optimum channel. \textcolor{black}{ The limited WL leads to the loss in precision of QF values, which results in multiple channels with identical QF values. In cases where more than one channel have maximum QF values, the channel selector unit selects the channel with a lower index as shown in Fig.~\ref{MABK5}.}

Next, we consider $K$ = 7, $N$ = 10000 and  two new distributions:\\
\noindent 1)\{$\mu_3, \sigma_3^2$\}  = \{(0.4,0.01), (0.45, 0.02), (0.35, 0.08), (0.33, 0.06), (0.37, 0.07), (0.46, 0.2), (0.38, 0.1)\} \\
\noindent 2) \{$\mu_4, \sigma_4^2$\}  =  \{(0.95, 0.03), (0.92,0.08), (0.88, 0.1), (0.87, 0.15), (0.9, 0.1), (0.98, 0.01), (0.82, 0.1)\}. 

\begin{figure}[!h]
\centering
\vspace{-0.3cm}
\subfloat[]{\includegraphics[width=0.475\textwidth]{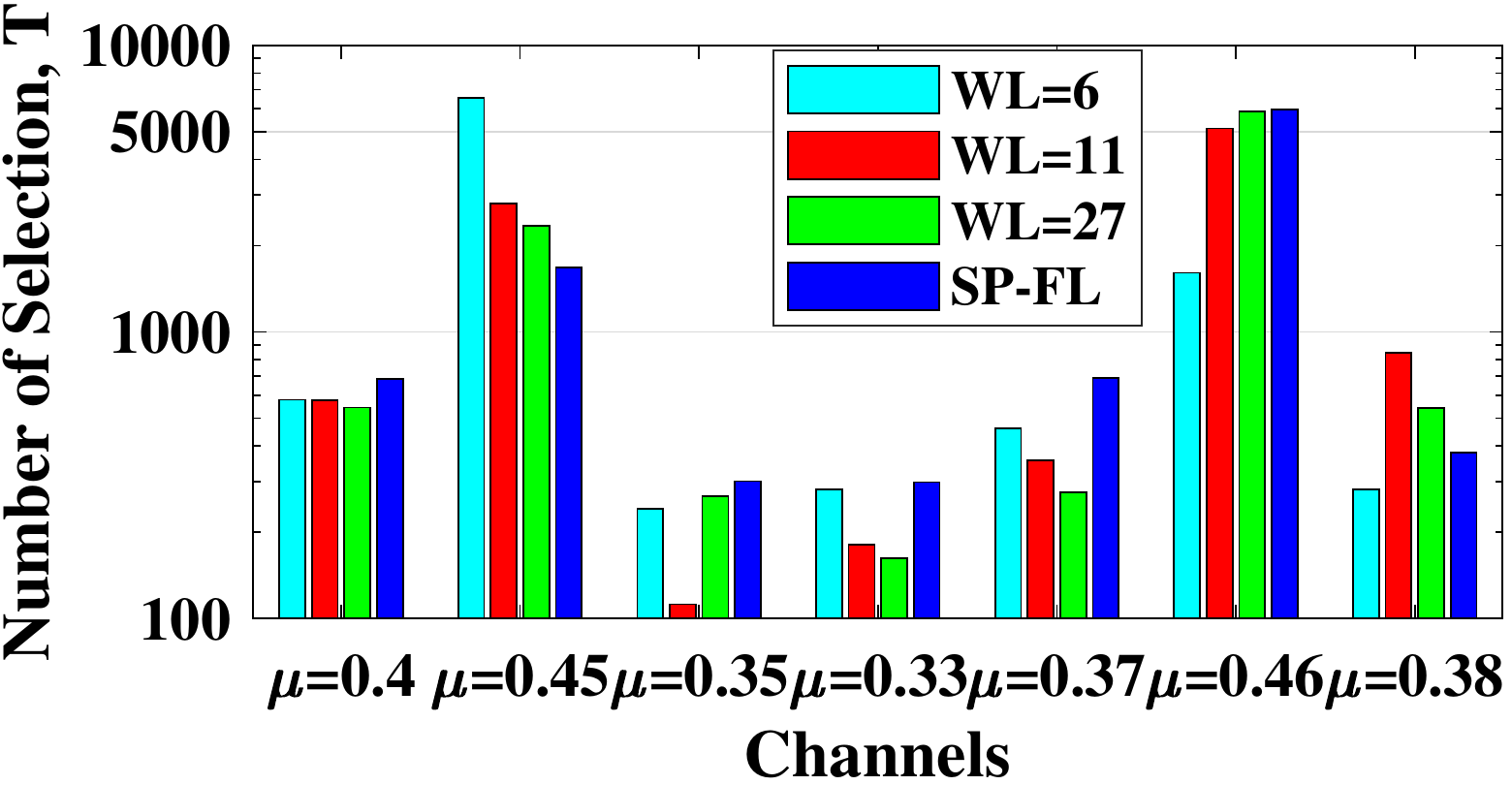}%
 \label{MABK71}}\\
 \vspace{-0.3cm}
\subfloat[]{\includegraphics[width=0.475\textwidth]{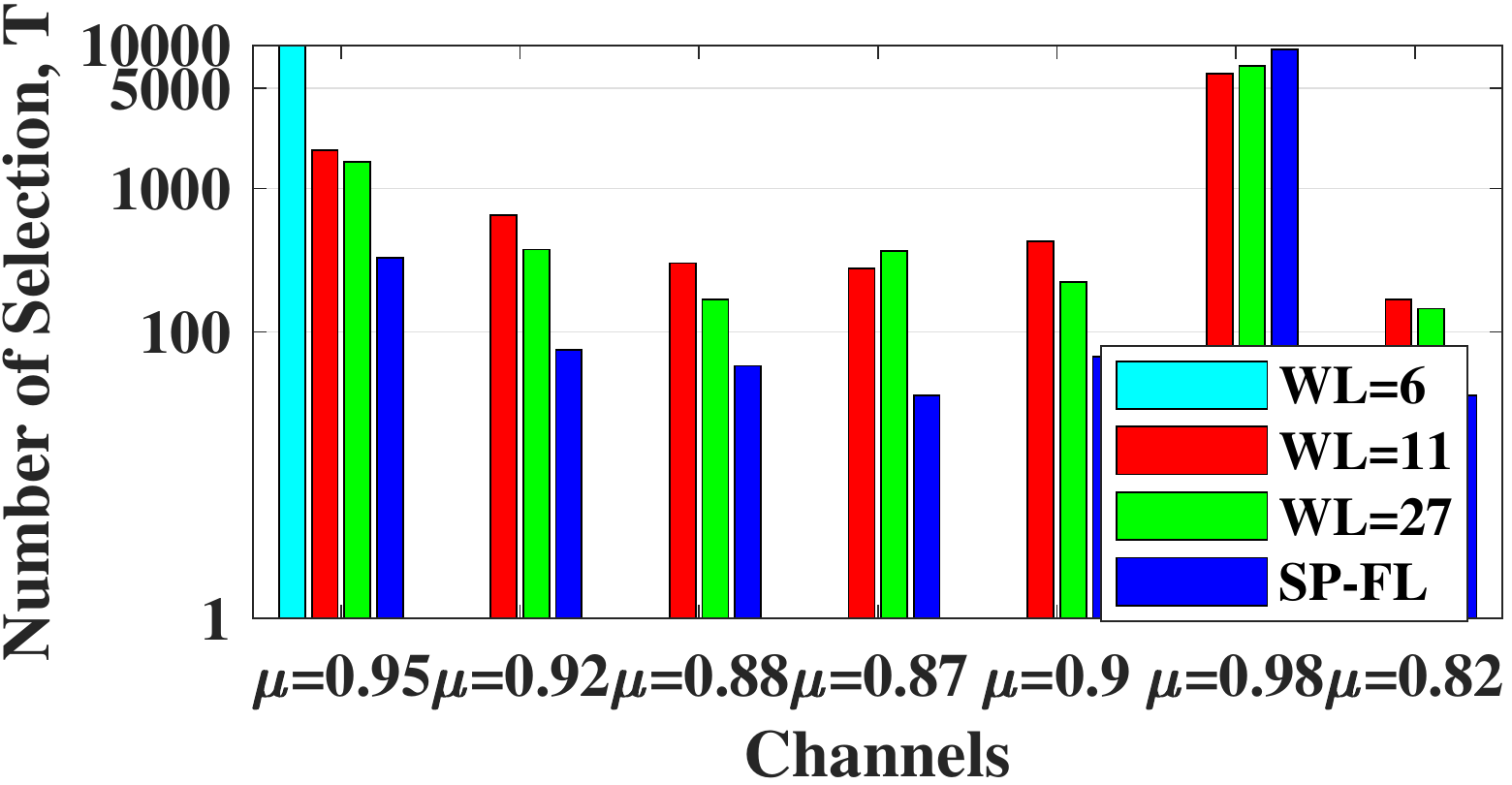}%
 \label{MABK72}}
\caption{Channel selection by UCB algorithms after $N=10000$ time slots with $K$ = 5, (a) $\mu_3$, and (b) $\mu_4$ }
\label{MABK7}
\vspace{-0.3cm}
\end{figure}

Here, the difference between channel statistics is small, which means learning and identifying the optimal channel is challenging. As shown in Fig.~\ref{MABK7}, proposed UCB architectures select the optimum channels a higher number of times except for WL=6. In both the distributions, UCB algorithms need to explore channels 2 and 6 in $\mu_3$ and channels 1 and 6 in $\mu_4$ the sufficient number of times before exploiting the optimal channel, i.e., channel 6 in $\mu_3$ and $\mu_4$. Similar to Fig.~\ref{MABK5}, the architecture with WL=6 fails to identify the optimal channel. For the rest of the discussion, we consider the MAB architecture with WL=11 as it offers the performance close to that of WL=27 and SP-FL architectures. Furthermore, it provides a significant reduction in resource utilization, as discussed later.

\vspace{-0.3cm}
\subsection{Performance Analysis of Wireless PHY}
Next, UCB architecture with WL=11 is integrated with wireless PHY architecture to compare the gain in performance due to MAB algorithms. Here, we consider $K=7$, $N=10000$, and six different channel statistics for performance evaluation in a different environment:\\
\noindent 1) $\mu_3$ (Same as Fig.~\ref{MABK71})\\
\noindent 2) $\mu_4$ (Same as Fig.~\ref{MABK72})\\
\noindent 3) $\mu_5$=\{0.3,0.3,0.35,0.4,0.45,0.5,0.55\}\\
\noindent 4) $\mu_6$=\{0.35 0.4 0.45 0.5 0.55 0.6 0.65\}\\
\noindent 5) $\mu_6$=\{0.45 0.5 0.55 0.6 0.65 0.7 0.75\} \\
\noindent 6) $\mu_6$=\{0.3 0.35 0.4 0.45 0.5 0.55 0.6\}

\begin{figure}[!h]
\centering
\includegraphics[width=0.475\textwidth]{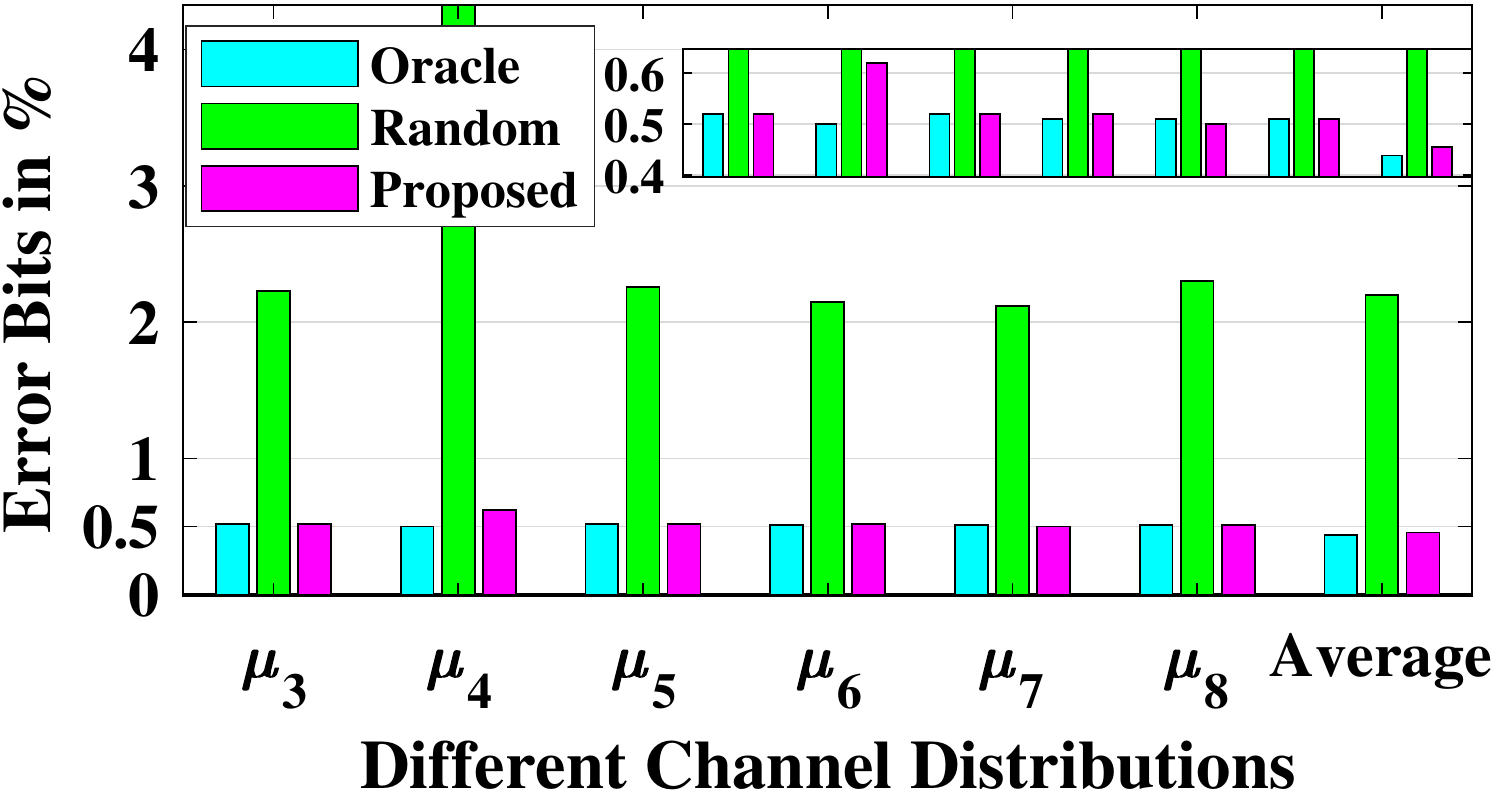}
\caption{Total received error bits in \% for different channel selection approaches.}
\vspace{-0.3cm}
\label{BER_chsel}
\end{figure}

In each time slot, 48 OFDM data symbols are transmitted, equivalent to 96 and 192 bits for QPSK and 16-QAM modulation. In Fig.~\ref{BER_chsel}, we compare the total number of bits received in error in \% for three different channel selection schemes: 1) Oracle scheme that always selects the optimal channel, 2) Random channel selection scheme, and 3) Proposed MAB based channel selection scheme. In each case, we have used a channel-based reconfigurable modulation scheme switching approach, i.e., when the channel is good, 16-QAM is used instead of QPSK. It can be observed that the proposed scheme offers significantly lower \% of error bits than a random channel selection approach, thereby validating the need for MAB based decision making for channel selection. Also, the performance of the oracle and the proposed approach is almost identical, thereby validating the optimality of the UCB algorithm as well as the correctness of its architecture to identify the optimal channel quickly.

\begin{figure}[!h]
\centering
\includegraphics[width=0.475\textwidth]{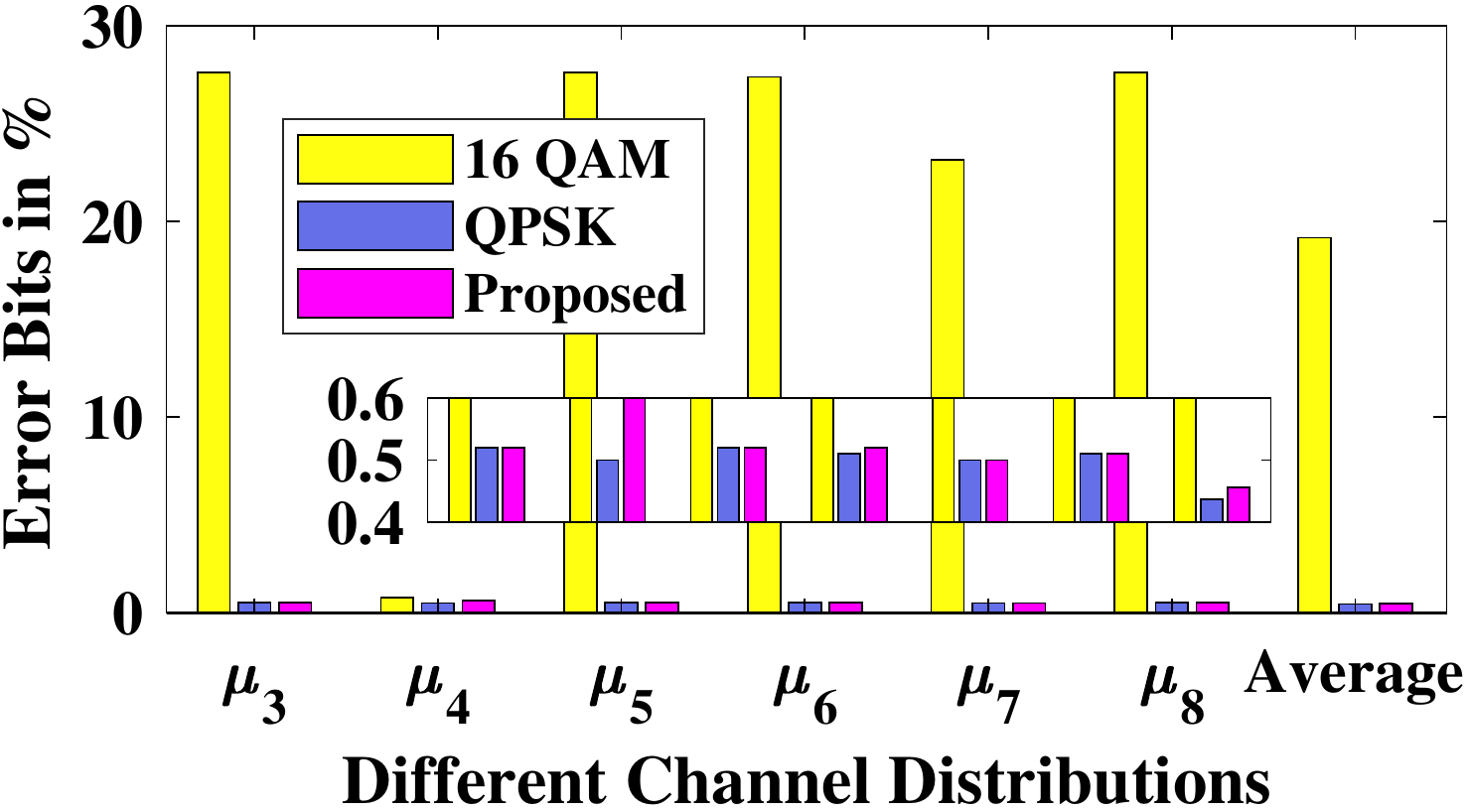}
\caption{Total received error bits in \% for different modulation scheme selection approaches.}
\label{BER_msel}
\end{figure}

Next, we repeat the above experiment to compare three different modulation selection approaches 1) Fixed 16-QAM, 2) Fixed QPSK, and 3) Proposed channel-based modulation switching scheme. Here, the channel selection is based on the MAB based approach. As shown in Fig.~\ref{BER_msel}, 16-QAM incurs a higher number of error bits due to performance degradation when the poor channel is selected. Note that around 25\% bits are received in error even though MAB selects the optimal channel a large number of times. It is obvious that the error rate will increase substantially for the random channel selection approach. As expected, the lower-order modulation scheme, such as QPSK, offers the lowest number of error bits. The proposed channel based modulation switching scheme provides significantly better performance than 16-QAM and around 0.5\% higher number of error bits than QPSK. \textcolor{black}{This penalty is due to the time taken by the MAB algorithm to estimate channel statistics (exploration time), which allows dynamic selection of appropriate modulation schemes for a given channel. Next, we discuss the effect of intelligent channel and modulation scheme selection on average throughput.}

As discussed before, PHY with 16-QAM transfers twice the number of bits than the PHY with QPSK in a given number of transmission slots. To take this into account, we compare the average throughput of the PHY for three different modulation selection approaches 1) Fixed 16-QAM, 2) Fixed QPSK, and 3) Proposed channel-based modulation switching scheme. Here, we skip the random channel selection approach due to the high error rate, as shown in Fig.~\ref{BER_chsel}. In Table~\ref{tab:thr}, we compare the total throughput in megabits per second (Mbps) and the total number of transmissions required to meet the BER of 1\%. This means that the 16-QAM and proposed architecture may need to re-transmission to meet given BER constraints. Here, the feedback time or the time interval between re-transmissions is assumed to 30\% of the transmission time, which means every re-transmission leads to a penalty in throughput. It can be observed that the proposed PHY offers throughput the same as that of QPSK based PHY when the channel quality is average or poor (All except the second column in Table~\ref{tab:thr}). On the other hand, when the channel quality is good (Second column in Table~\ref{tab:thr}), the proposed PHY offers significantly better throughput than other PHY. For six different channel distributions, out of which five are average or poor and one is good, proposed PHY provides better throughput and fewer re-transmission.  Though the error rate of QPSK based PHY is lower than the proposed PHY in Fig.~\ref{BER_msel}, the latter offers higher throughput than the former. For the environment with excellent or above-average channel conditions, the gain in throughput over QPSK based PHY will further increase. 

\begin{table}[!h]
\centering
\caption{Comparison of throughput in Mbps and Number of re-transmission for different channel distributions}
\label{tab:thr}
\renewcommand{\arraystretch}{1.3}
\resizebox{0.5\textwidth}{!}{%
\begin{tabular}{|l|l|l|l|l|l|l|l|}
\hline
 & \textbf{$\mu_3$} & \textbf{$\mu_4$} & \textbf{$\mu_5$} & \textbf{$\mu_6$} & \textbf{$\mu_7$} & \textbf{$\mu_8$} & \textbf{Average} \\ \hline
\textbf{QPSK} & 50 (1) & 50 (1) & 50 (1) & 50 (1) & 50 (1) & 50 (1) & 50 (1) \\ \hline
\textbf{QAM} & 38.7 (5) & 76.5 (2) & 38.7 (5) & 38.8 (5) & 45.5 (4) & 38.7 (5) & 43.4  (4.3) \\ \hline
\textbf{Proposed} & 50 (1) & \textbf{90 (1)} & 50 (1) & 50 (1) & 50 (1) & 50 (1) & \textbf{53 (1)} \\ \hline
\end{tabular}%
}
\end{table}

The fewer number of re-transmissions makes the proposed PHY power-efficient leading to longer battery life. Also, since the error rate of 16-QAM is significantly high (above 24\%), it is not suitable for a low coding rate scenario since the entire frame needs to be transmitted again due to lower code correction capability. On the other hand, the proposed approach exploits the high throughput advantages of 16-QAM and offers higher resilience to block errors, mainly due to the channel-based modulation selection approach.
\vspace{-0.3cm}
\subsection{Complexity Comparison}
In this section, we present the detailed complexity comparison of the proposed architecture. First, we consider the proposed architecture with a reconfigurable channel based modulation scheme and fixed MAB algorithm, i.e., only the UCB algorithm is realized with a fixed $K$. As shown in Table~\ref{resMod}, we consider the Velcro approach where QPSK and QAM modulator and demodulator blocks are realized in parallel, i.e., all four blocks are always active. In the proposed approach using DPR, the appropriate modulation scheme is dynamically configured on-the-fly. As expected, the proposed approach needs slightly lower LUTs and half the number of flip-flops (FFs) than the Velcro approach. Since the QPSK block transmits 2 bits in one resource element compare to four bits in QAM, we need a separate clock network for QPSK and QAM. Thus, the proposed approach needs fewer FFs since only one clock is active compared to two clock networks in the Velcro approach. For fixed-point implementation with WL= 27, 40\%, and 80\% reduction in LUTs and DSP48s (embedded multiplier and accumulator units) with almost identical BER and throughput performance as that of SP-FL architecture. With WL=11, further improvement in savings of up to 60\% and 87\% in resources can be achieved with negligible degradation in the BER and throughput. With the increase in the number of supported modulation schemes and the addition of other reconfigurable modules such as channel coding and waveform modulation, the proposed approach can offer further improvement in resource utilization.

\begin{table}[!h]
\caption{Resource utilization comparison with reconfigurable modulation and fixed $K=5$}
\renewcommand{\arraystretch}{1.2}
\label{resMod}
\resizebox{0.5\textwidth}{!}{%
\begin{tabular}{|c|c|c|c|c|}
\hline
\textbf{\begin{tabular}[c]{@{}c@{}}Module\end{tabular}}              & \textbf{Approach}                                                            & \textbf{LUTs}  & \textbf{FFs}  & \textbf{DSPs} \\ \hline
\multirow{4}{*}{\begin{tabular}[c]{@{}c@{}}Wireless PHY with \\ QAM-16/QPSK via DPR \\ and $K=5$\end{tabular}} & \textbf{\begin{tabular}[c]{@{}c@{}}Velcro\\(SP-FL)\end{tabular}} & 64828 & 41175 & 342  \\ \cline{2-5} 
                                                                                                & \textbf{\begin{tabular}[c]{@{}c@{}}Proposed\\  (SP-FL)\end{tabular}}      & \textbf{64796} & \textbf{21042} & \textbf{342}   \\ \cline{2-5} 
                                                                                                & \textbf{\begin{tabular}[c]{@{}c@{}}Proposed \\  WL=27\end{tabular}}       & \textbf{\begin{tabular}[c]{@{}c@{}}\textbf{39198}\\ (-40\%)\end{tabular}}   & \begin{tabular}[c]{@{}c@{}}30390 \\  (+44\%)\end{tabular}   & \textbf{\begin{tabular}[c]{@{}c@{}}\textbf{67}\\ (-80\%)\end{tabular}}   \\ \cline{2-5} 
                 
                                                                                                 & \textbf{\begin{tabular}[c]{@{}c@{}}Proposed \\ WL=11\end{tabular}}       & \textbf{\begin{tabular}[c]{@{}c@{}}\textbf{26808}\\(-60\%)\end{tabular}}
& \begin{tabular}[c]{@{}c@{}}26300 \\  (+24\%)\end{tabular} & \textbf{\begin{tabular}[c]{@{}c@{}}\textbf{47}\\ (-87\%)\end{tabular}}  \\  \hline
\end{tabular}
}
\end{table}

\begin{table*}[!t]
\centering
\caption{Complexity comparison of various configuration in wireless PHY architecture}
\renewcommand{\arraystretch}{1.4}
\label{tab:complexity}
\resizebox{\textwidth}{!}{%
\begin{tabular}{|c|l|c|c|c|c|c|c|c|c|c|c|c|c|c|}
\hline
\multirow{2}{*}{No.} & \multicolumn{1}{c|}{\multirow{2}{*}{\textbf{Architecture}}} & \multirow{2}{*}{\textbf{Approach}} & \multicolumn{3}{c|}{\textbf{No. of LUTs}} & \multicolumn{3}{c|}{\textbf{No. of FFs}} & \multicolumn{3}{c|}{\textbf{No. of DSP48}} & \multicolumn{3}{c|}{\textbf{Dynamic Power (W)}} \\ \cline{4-15} 
 & \multicolumn{1}{c|}{} &  & \textbf{K=3} & \textbf{K=4} & \textbf{K=5} & \textbf{K=3} & \textbf{K=4} & \textbf{K=5} & \textbf{K=3} & \textbf{K=4} & \textbf{K=5} & \textbf{K=3} & \textbf{K=4} & \textbf{K=5} \\ \hline
\multirow{2}{*}{1} & \multirow{2}{*}{\textbf{\begin{tabular}[c]{@{}l@{}}Wireless PHY with QPSK/QAM in Velcro,\\ fixed UCB algorithm,  tunable $K \leq K_{max}$\end{tabular}}} & \textbf{Velcro} & \multicolumn{3}{c|}{24173} & \multicolumn{3}{c|}{19245} & \multicolumn{3}{c|}{127} & \multicolumn{3}{c|}{0.41} \\ \cline{3-15} 
 &  & \textbf{Proposed} & 17187 & 20680 & 24173 & 15407 & 17326 & 19245 & 87 & 107 & 127 & 0.36 & 0.38 & 0.41 \\
 &  &  & \textbf{-29\%} & \textbf{-15\% }&  & \textbf{-20\% } & \textbf{-10\%} &  &  \textbf{-32\% } & \textbf{-16\% } &  & \textbf{-12.2\% } & \textbf{-7.3\% } &  \\
 \hline
 \hline
\multirow{2}{*}{2} & \multirow{2}{*}{\textbf{\begin{tabular}[c]{@{}l@{}}Wireless PHY with QPSK/QAM in Velcro,\\ fixed UCB\_V algorithm, tunable $K \leq K_{max}$\end{tabular}}} & \textbf{Velcro} & \multicolumn{3}{c|}{25263} & \multicolumn{3}{c|}{20110} & \multicolumn{3}{c|}{142} & \multicolumn{3}{c|}{0.46} \\ \cline{3-15} 
 &  & \textbf{Proposed} & 17841 & 21552 & 25263 & 15926 & 18018 & 20110 & 96 & 119 & 142 & 0.39 & 0.42 & 0.46 \\ 
 &  & & \textbf{-30\%} & \textbf{-15\%} &  & \textbf{-21\%} & \textbf{-11\%} &  & \textbf{-33\%} & \textbf{-17\%} & & \textbf{-15.2\%} & \textbf{-8.7\%} &  \\ 
 \hline
  \hline
 
\multirow{2}{*}{3} & \multirow{2}{*}{\textbf{\begin{tabular}[c]{@{}l@{}}Wireless PHY with QPSK/QAM in Velcro,\\ fixed UCB\_T algorithm,  tunable $K \leq K_{max}$\end{tabular}}} & \textbf{Velcro} & \multicolumn{3}{c|}{28808} & \multicolumn{3}{c|}{21120} & \multicolumn{3}{c|}{127} & \multicolumn{3}{c|}{0.46} \\ \cline{3-15} 
 &  & \textbf{Proposed} & 19968 & 24388 & 28808 & 16538 & 18826 & 21120 & 87 & 107 & 127 & 0.39 & 0.42 & 0.46 \\
 &  &  & \textbf{-31\%} & \textbf{-16\%} &  & \textbf{-22\%} & \textbf{-11\%} &  & \textbf{-32\%} & \textbf{-16\%} & & \textbf{-15.2\%} & \textbf{-8.7\%} & \\
 \hline\hline
\multirow{4}{*}{4} & 

\multirow{4}{*}{\textbf{\begin{tabular}[c]{@{}l@{}}Wireless PHY with \\ QPSK/QAM in Velcro,\\ reconfigurable MAB and \\ tunable $K \leq K_{max}$\end{tabular}}} & \textbf{Velcro} & \multicolumn{3}{c|}{64828} & \multicolumn{3}{c|}{41175} & \multicolumn{3}{c|}{342} & \multicolumn{3}{c|}{0.75} \\ \cline{3-15} 
 &  &\textbf{Proposed} & 19968 & 24388 & 28808 & 16538 & 18826 & 21120 & 96 & 119 & 142 & 0.55 & 0.65 & 0.75 \\ 
 
  &  & \textbf{ SP-FL} & \textbf{-70\%} & \textbf{-62\%} & \textbf{-56\%} & \textbf{-75\%} & \textbf{-55\%} & \textbf{-49\%} & \textbf{-72\%} & \textbf{-65\%} & \textbf{-59\%} & \textbf{-26.7\%} & \textbf{-13.3\%} &  \\ 
  
 \cline{3-15} 
 &  & \textbf{Proposed} & 15326 & 18255 & 21184 & 14889 & 16768 & 18647 & 39 & 43 & 47 & 0.53 & 0.62 & 0.7 \\ 
 &  & \textbf{WL=27} & \textbf{-77\%} & \textbf{-72\%} & \textbf{-68\%} & \textbf{-64\%} & \textbf{-60\%} & \textbf{-55\%} & \textbf{-89} & \textbf{-87\%} & \textbf{-87\%} & \textbf{-29.3\%} & \textbf{-17.3\%} & \textbf{-0.03\%} \\ 
 
 \cline{3-15} 
 &  & \textbf{Proposed} & 11887 & 13650 & 15413 & 14091 & 15678 & 17265 & 33 & 35 & 37 & 0.35 & 0.37 & 0.39 \\
 &  & \textbf{WL=11} & \textbf{-82\%} & \textbf{-79\%} & \textbf{-77\%} & \textbf{-66\%} & \textbf{-62\%} & \textbf{-59\%} & \textbf{-91\%} & \textbf{-90\%} & \textbf{-90\%} & \textbf{-53.3\%} & \textbf{-50.7\%} & \textbf{-48\%} \\ 
  \hline
 \hline
\multirow{4}{*}{5} 
& \multirow{4}{*}{\textbf{\begin{tabular}[c]{@{}l@{}}Complete reconfigurable Wireless PHY with \\ tunable $K \leq K_{max}$\end{tabular}}} & \textbf{Velcro} & \multicolumn{3}{c|}{64828} & \multicolumn{3}{c|}{41175} & \multicolumn{3}{c|}{342} & \multicolumn{3}{c|}{0.75} \\ \cline{3-15} 
 &  & \textbf{Proposed} & 19909 & 24329 & 28749 & 16460 & 18748 & 21002 & 96 & 119 & 142 & 0.55 & 0.65 & 0.75 \\ 
  &  & \textbf{SP-FL} & \textbf{-70\%} & \textbf{-63\%} & \textbf{-56\%} & \textbf{-60\%} & \textbf{-55\%} & \textbf{-49\%} & \textbf{-72\%} & \textbf{-65\%} & \textbf{-59\%} & \textbf{-27\%} & \textbf{-14\%} &  \\ 
  \cline{3-15} 
 &  & \textbf{Proposed} & 15267 & 18196 & 21125 & 14811 & 16690 & 18569 & 39 & 43 & 47 & 0.53 & 0.62 & 0.7 \\
  &  & \textbf{WL=27} & \textbf{-77\%} & \textbf{-72\%} & \textbf{-68\%} & \textbf{-64\%} & \textbf{-60\%} & \textbf{-55\%} & \textbf{-89\%} & \textbf{-87\%} & \textbf{-87\%} & \textbf{-29.3\%} & \textbf{-17.3\%} & \textbf{-7\%} \\ 
 \cline{3-15} 
 &  & \textbf{Proposed} & 11828 & 13591 & 15354 & 14013 & 15600 & 17187 & 33 & 35 & 37 & 0.35 & 0.37 & 0.39 \\
  &  & \textbf{WL=11} & \textbf{-82\%} & \textbf{-79\%} & \textbf{-77\%} & \textbf{-66\%} & \textbf{-62\%} & \textbf{-59\%} & \textbf{-91\%} & \textbf{-90\%} & \textbf{-90\%} & \textbf{-53.3\%} & \textbf{-50.7\%} & \textbf{-48\%} \\ 
 \hline
 \hline
6
& \textbf{Reconfigurable Wireless PHY on PS} & \textbf{ARM} & \multicolumn{3}{c|}{6026} & \multicolumn{3}{c|}{8533} & \multicolumn{3}{c|}{27} & \multicolumn{3}{c|}{0.24} \\ \hline
\end{tabular}%
}
\end{table*}

Next, we extend the architecture with reconfigurability at the MAB algorithm (i.e., architecture which can dynamically switch between UCB, UCB\_T, and UCB\_V algorithms) and $K$ (i.e., number of channels can be tuned to any values less than or equal to $K_{max}$. In Table~\ref{tab:complexity}, we consider six different architectures which are discussed below:
\begin{enumerate}
 \item First architecture implements the modulation scheme via the Velcro approach, supports only the UCB algorithm and offers the DPR at the channel level, i.e., reconfigurable $K$. 
 \item Second architecture is similar to the first architecture except that the UCB algorithm is replaced with the UCB\_V algorithm. 
 \item Third architecture is similar to the first architecture except that the UCB algorithm is replaced with the UCB\_T algorithm. 
 \item Forth architecture implements the modulation scheme via the Velcro approach with reconfigurable MAB and $K$, i.e., architecture can dynamically switch between UCB, UCB\_V and UCB\_T algorithms in addition to any $K \leq K_{max}$.
 \item Fifth architecture is entirely reconfigurable and can dynamically select the modulation scheme, MAB algorithm, as well as $K$.
 \item Six architecture consists of PHY on PL and MAB algorithm on PS. 
 \end{enumerate}

In the first row of Table~\ref{tab:complexity}, we compare the first architecture realized using the Velcro approach and proposed approach via DPR. It can be observed that the resource utilization of the proposed approach depends on the active number of channels, $K$ compared to the Velcro approach, which corresponds to architecture with $K=K_{max}$, i.e., all blocks are active all the time compared to dynamic activation and deactivation of channels in the proposed approach. In the next two rows, we compare the architectures by replacing UCB with UCB\_V and UCB\_T algorithms. Since QF calculation in UCB\_T (Eq.~\ref{qf_ucbt}) and UCB\_V (Eq.~\ref{qf_ucbv}) are computationally complex than UCB algorithms (Eq.~\ref{qf_ucb}), second and third architectures have higher resource utilization than first architecture. In all the three architectures, proposed architectures offer 7-15\% savings in dynamic power consumption of the reconfigurable region over the Velcro approach for $K< K_{max}$.

In the fourth architecture, reconfigurability is added at the MAB level in addition to $K$. It can be observed that the proposed approach needs 56-70\% lower LUTs, 55-65\% lower FFs, and 59-72\% fewer DSP48 units. Also, the proposed architecture offers power savings up to 53\% in the reconfigurable region. We have also realized the fixed-point configuration of the proposed architecture for different WL. As discussed above, WL of 27 offers the BER and throughput performance similar to that of SP-FL architecture, while WL of 11 is the minimum WL, after which the BER and throughput performance degrade significantly. Thus, the selection of appropriate WL is essential to achieve the right balance between performance and complexity. 

In the fifth architecture, we have included the reconfigurability at the modulation block as well, and it can be observed a slight improvement in resource utilization similar to Table~\ref{resMod}. With the increase in the number of modulation schemes, the proposed architecture will further offer savings in resources than the Velcro approach. In the sixth architecture, we have explored the ARM processor along with co-processor NEON. As expected, it provides the lowest resource utilization and power consumption. However, overall latency is significantly high due to sequential implementation in PS, which affects the throughput.

In Table~\ref{latencyMAB}, we compare the latency of the fifth architecture for three configurations: 1) MAB algorithm in PL, 2) MAB algorithm in the ARM processor of PS, 3) MAB algorithm in ARM+NEON processor of PS. The PHY is intentionally realized in PL due to the practical constraints of interfacing the output of the transmit PHY with antenna via digital and analog front-end. Such an interface is available only on PL due to a large number of interface pins. The latency corresponds to the transmission of 10000 OFDM frames, each consisting of 320 bits of user data (4 OFDM symbols). It can be observed that the PL implementation of MAB offers the lowest latency, and performance improves as $K$ increases. This is due to the parallel execution of the QF function in PL compared to sequential PS execution. \textcolor{black}{From a numerical perspective, OFDM symbol time in the proposed architecture is 17.5$\mu$s. In 5G with sub-carrier spacing of 15 kHz and 30 kHz, the desired OFDM symbol time is 71.4$\mu$s and 35.7$\mu$s, respectively. Thus, the proposed architecture on ZSoC meets the 5G symbol time requirement. On the other hand, symbol time in the case of ARM and ARM+NEON platforms depends on the value of $K$. For a reasonable value of $K \geq 20$, it is incredibly challenging to meet the 5G latency requirement. Thus, in real networks with $K \geq 20$, the proposed architecture can significantly improve execution time, thereby supporting high throughput and low latency applications.}

\begin{table}[!h]
\caption{Execution time of different architectures}
\renewcommand{\arraystretch}{1.3}
\label{latencyMAB}
\resizebox{0.5\textwidth}{!}{%
\begin{tabular}{|c|c|c|c|c|c|c|c|c|c|c|}
\hline
\multirow{2}{*}{\textbf{Algorithm}} & \multirow{2}{*}{\textbf{Approach}} & \multirow{2}{*}{\textbf{\begin{tabular}[c]{@{}c@{}}ZSoC\\ (in ms)\end{tabular}}} & \multicolumn{4}{c|}{\textbf{\begin{tabular}[c]{@{}c@{}}PS (ARM)\\ (in ms)\end{tabular}}}          & \multicolumn{4}{c|}{\textbf{\begin{tabular}[c]{@{}c@{}}PS (ARM + NEON)\\ (in ms)\end{tabular}}}   \\ \cline{4-11} 
                                    &                                    &                                                                                  & \textbf{K=2}           & \textbf{K=3}           & \textbf{K=4}           & \textbf{K=5}           & \textbf{K=2}           & \textbf{K=3}           & \textbf{K=4}           & \textbf{K=5}           \\ \hline
\multirow{3}{*}{Architecture 5}             & SP-FL                        & \textbf{700}                                                                  & \multirow{3}{*}{719} & \multirow{3}{*}{729} & \multirow{3}{*}{737} & \multirow{3}{*}{747} & \multirow{3}{*}{713} & \multirow{3}{*}{720} & \multirow{3}{*}{726} & \multirow{3}{*}{733} \\ \cline{2-3}
                                    & WL=27                              & \textbf{700}                                                                  &                        &                        &                        &                        &                        &                        &                        &                        \\ \cline{2-3}
                                    & WL=11                              & \textbf{700}                                                                  &                        &                        &                        &                        &                        &                        &                        &                        \\ \hline
\end{tabular}
}
\end{table}

\section{Conclusions and Future Directions}
\label{sec:con}
In this paper, we present the novel design and implementation of an intelligent and reconfigurable physical layer (PHY) for next-generation wireless transceivers. The proposed PHY is integrated with a multi-armed bandit (MAB) based learning algorithm to learn and identify the optimum channel in any given environment. To the best of our knowledge, the proposed work is the first-ever implementation of the MAB algorithm on SoC. \textcolor{black}{The MAB algorithm offers intelligence to dynamic partial reconfiguration (DPR) based reconfigurable PHY, and detailed performance analysis shows that the proposed architecture offers 3 Mbps higher average throughput, lower resource utilization (overall savings of 50-80\% in LUTs, 49-60\% in FFs, and 60-90\% in DSP units), and 14-50\% lower power consumption compared to conventional PHY.} The proposed demonstration of the feasibility of intelligent and reconfigurable PHY opens up a wide range of research directions involving the integration of PHY and learning algorithms. In the future, we will explore various other learning algorithms, along with their synthesizable low complexity architectures. In addition, we aim to develop an intelligent and reconfigurable prototype for 3GPP 5G PHY along with performance analysis in the real-radio environment and feasibility for real-world applications. 


\section*{Acknowledgment}
We would like to thank the authors of the paper \cite{2017_ROFDM_pham} for source codes used in the implementation of wireless PHY.

\bibliographystyle{IEEEtran}
\bibliography{biblio}

\begin{IEEEbiography}[{\includegraphics[width=1in,height=1.25in,clip,keepaspectratio]{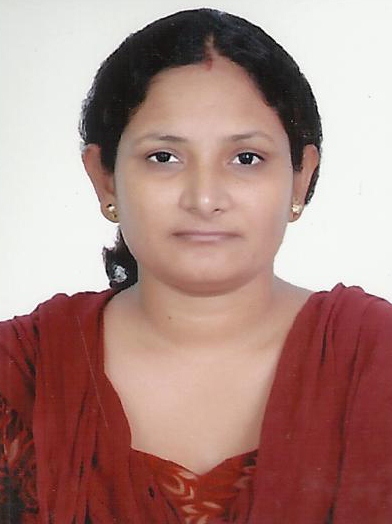}}]%
{Neelam}
received the M. Tech degree in Signal Processing from GGSIP Delhi .She is currently pursuing the PhD degree in Reconfigurable Radio on SOC at IIIT, Delhi. She is having more than of 10 years  of industrial experience in Board designing.
Her current research interest include designing and mapping of reconfigurable and intelligent wireless PHY on Hardware.
\end{IEEEbiography}

\begin{IEEEbiography}[{\includegraphics[width=1in,height=1.25in,clip,keepaspectratio]{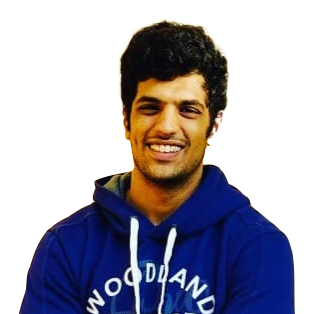}}]%
{S. V. Sai Santosh}
is an undergraduate student in Electronics and Communication Engineering at Indraprastha Institute of Information Technology Delhi, India.
His current research interests include the design of efficient algorithms and hardware for reconfigurable and intelligent wireless and AI applications. 
\end{IEEEbiography}

\begin{IEEEbiography}[{\includegraphics[width=1in,height=1.25in,clip,keepaspectratio]{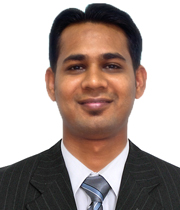}}]%
	{Sumit J. Darak}
	received his undergraduate degree in Electronics and Telecommunications Engineering from Pune University, India in 2007, and PhD degree from the School of Computer Engineering, Nanyang Technological University (NTU), Singapore in 2013.
	He is currently an Associate Professor at Indraprastha Institute of Information Technology, Delhi, (IIIT Delhi) India. Before joining IIIT Delhi, he worked as postdoctoral researcher at the CominLabs Excellence Center, CentraleSupélec, France from March 2013 to November 2014. He is the recipient of the \textit{DST Inspire Faculty Award} (2015-2020), \textit{Best Demo Award} at CROWNCOM 2016, \textit{Second Best Paper Award} at IEEE DASC 2017, \textit{Young Scientist Paper Award} at URSI 2014 and 2017, \textit{Second Best Poster Award} at COMSNETS 2019 and \textit{Best Paper Award} at NCSIPA 2009. He is a recipient of National Instruments (NI) academic research grant (2017, 2018) and Core Research Grant from DST-SERB, India. Dr. Sumit is also working as 5G consultant for VVDN Technologies since March 2019. His current research interests include the design of efficient algorithms and mapping to reconfigurable and intelligent architectures for wireless and AI applications. 
	
\end{IEEEbiography}

  

\section*{Supplementary material}

# Tutorial: Intelligent & Reconfigurable Wireless Physical Layer (PHY)

## Introduction

In this lab, you will use Vivado IPI and Software Development Kit to create a reconfigurable peripheral using ARM Cortex-A9 processor system on Zynq. You will use Vivado IPI to create a top-level design, which includes the Zynq processor system as a sub-module. During the PR flow, you will define four Reconfigurable Partitions (MAB, PHY) having Reconfigurable Modules (UCB, UCB_V, UCB_T), (16-QAM, QPSK) respectively. You will create multiple Configurations and run the Partial Reconfiguration implementation flow to generate full and partial bitstreams. You will use ZC706 to verify the design in hardware using a SD card to initially configure the FPGA, and then partially reconfigure the device using the PCAP under user software control.

**GitHub Link:** https://github.com/Sai-Santosh-99/ReconfigPHY

## Design Description

The purpose of this lab exercise is to implement a design that can be intelligent & dynamically reconfigurable Wireless Physical Layer using PCAP resource and PS sub-system. The system consists of four Online Machine Learning peripherals (for 4 channels) having three unique function calculation capabilities (UCB_T, UCB_V, UCB), one Wireless Transmitter PHY having two modulation capabilities (16-QAM, QPSK), one Wireless Receiver PHY having two demodulation capabilities (for 16-QAM, for QPSK) and a reconfigurable Comparator design. The architecture is designed such that the No. of channels as well as the OML algorithm can be reconfigured in the MAB block on-the-fly. Also, the modulation & demodulation blocks inside the Transmitter PHY & Receiver PHY respectively can also be configured on-the-fly based on the channel characteristics learnt by the MAB block. The Transmitter PHY selects the channel for transmission based on the input it receives from the MAB block. The Transmitter PHY then transmits the bits to the receiver where the pilot power ration is calculated. This ratio is then passed to the MAB block using the AXI-Lite interface. The MAB block then uses this ratio as the reward for the particular channel used to transmit the data. It then calculates the Q-Factor for all channels and then selects the channel with the maximum value to be used for transmission in the next window. The user verifies the functionality using a user application. The dynamic modules are reconfigured using the PCAP resource available through Device Configuration block. The design is shown in Figure 1.

## General Flow for this Lab

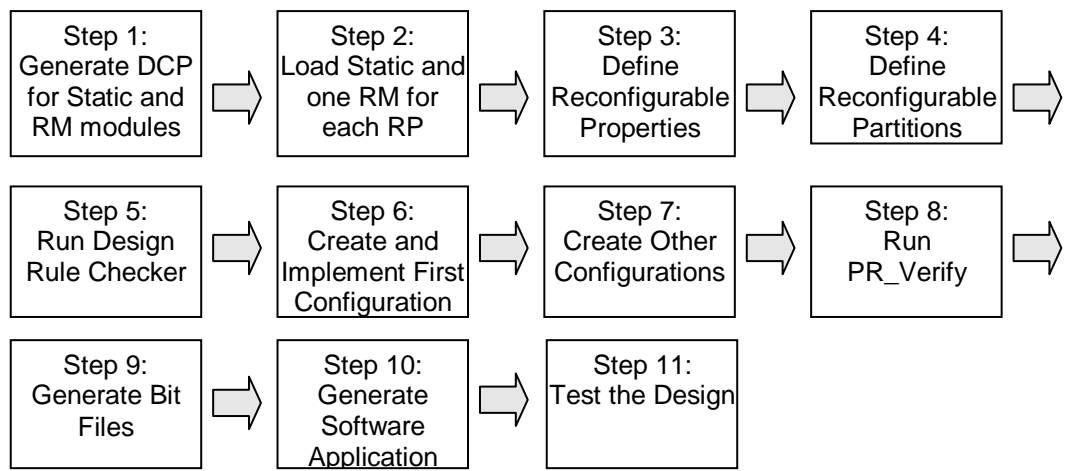

## Design

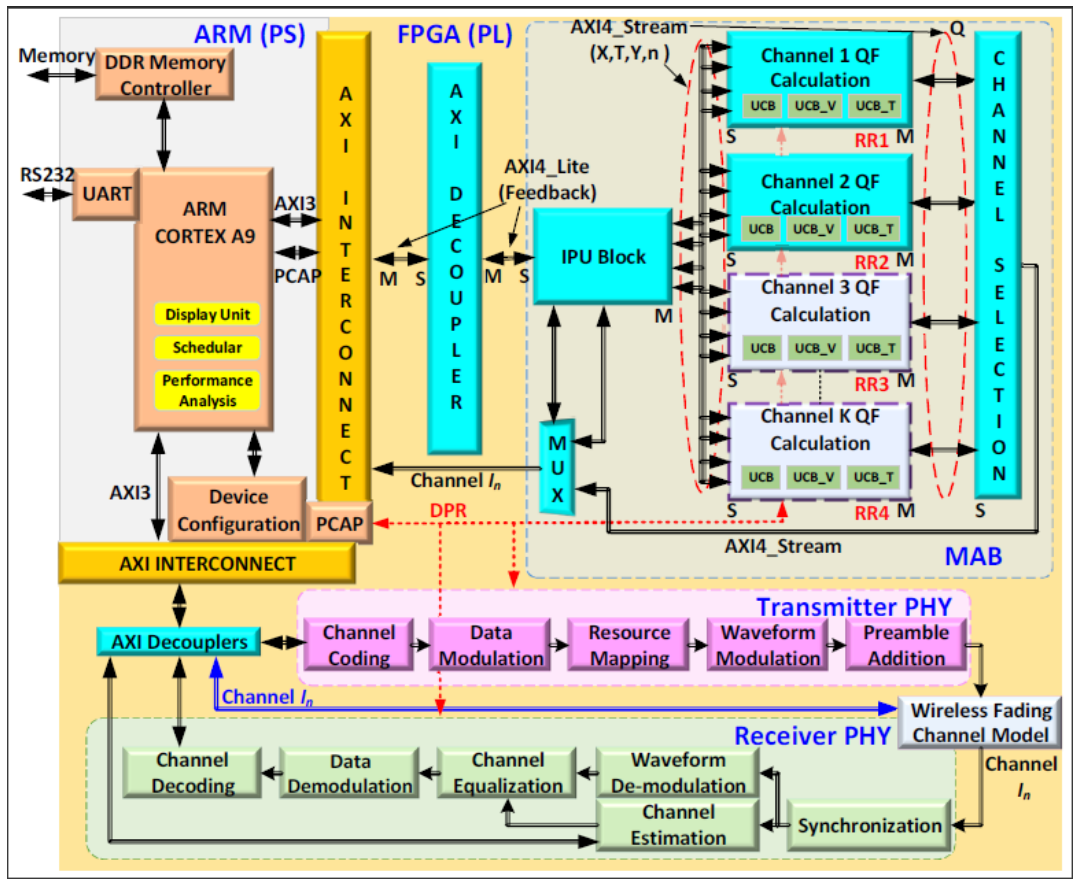

**Figure 1. The design**

## Procedure

This lab is separated into steps that consist of general overview statements that provide information on the detailed instructions that follow. Follow these detailed instructions to progress through the lab.

## Generate DCPs for the Static Design and RM Modules        Step 1

**1-1.    Start Vivado and execute the provided Tcl script to create the design check point for the static design having one RP.**

**1-1-1.**   Open **Vivado** by selecting **Start > All Programs > Xilinx Design Tools > Vivado 2018.2 > Vivado 2018.2**

**1-1-2.**   In the Tcl Shell window enter the following command to change to the lab directory and hit **Enter**.

```
cd c:/Summer/Tutorial
```

**1-1-3.**   Generate the PS design executing the provided Tcl script.

```
source blockDesign.tcl
```

This script will create the block design called system, instantiate ZYNQ PS with SD 0 and UART 1 interfaces enabled. It will also enable the GP0 interface along with FCLK0 and RESET0_N ports. The provided OFDM IP, MAB IPs, TRANSFER IPs, & COMPARE IPs will also be

instantiated. It will then create a top-level wrapper file called design_1_wrapper.v which instantiates the design_1.bd (the block design).

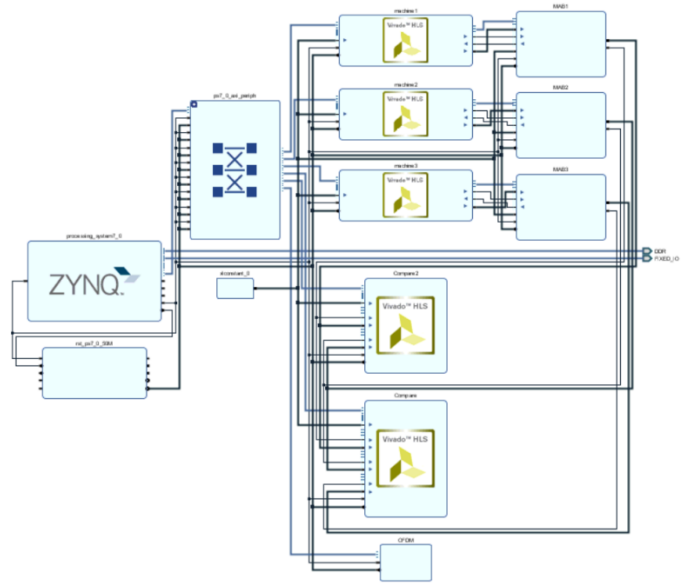

**Figure 2. The system block design**

**1-1-4.** Select the **Address Editor** tab. Expand **Data**. Expand **Unmapped Slaves**, if any, and right-click and select **Assign Address.**

**1-1-5.** Select **Tools** > **Validate Design.**

**1-1-6.** Select **File** > **Save Block Design.**

## 1-2. Synthesize the design to generate the dcp for the static logic of the design.

**1-2-1.** Click **Run Synthesis** under the *Synthesis* group in the *Flow Navigator* to run the synthesis process.

Wait for the synthesis to complete. When done click **Cancel**.

**1-2-2.** Using the windows explorer, copy the **design_1_wrapper.dcp** file from *tutorial\tutorial.runs\synth_*1 into the *Synth\Static* directory under the current lab directory.

**1-2-3.** Copy design checkpoints for the auto_pc, informTransfer_0, informTransfer_1, informTransfer_2, streamBlank_0, streamBlank_1, streamBlank_2, DPR_QAM_QPSK_0, compare_0, comparePR_0, xbar_0, rst_ps7_0_50M, and processing_system7_0 instances to *Synth\Static* to sit alongside system_wrapper.dcp

**1-2-4.** Close the project by typing the `close_project` command in the Tcl console or selecting **File > Close Project**.

## Load Static and one RM for the RP in Vivado                                        Step 2

**Since all required netlist files (dcp) for the design are already given in the Synth folder, you will use Vivado to floorplan the design, define Reconfigurable Partitions, add Reconfigurable Modules, run the implementation tools, and generate the full and partial bitstreams.**

**2-1.** In this step you will load the static and one RM designs for the RP.

**2-1-1.** In the Tcl Shell window enter the following command to change to the lab directory and hit **Enter**.

```
cd c:/Summer/Tutorial
```

**2-1-2.** Execute the following Tcl script to load the static design checkpoint.

```
source load_design_checkpoints.tcl
```

The script will do the following:

- Load the static design using the **open_checkpoint** command.

    ```
    open_checkpoint Synth/Static/design_1_wrapper.dcp
    ```

- Load the IP checkpoint for the Processing System by using the **read_checkpoint** command.

    ```
    read_checkpoint -cell design_1_i/processing_system7_0
    Synth/Static/system_processing_system7_0_0.dcp
    ```

- Load the IP checkpoint for the Processing Reset by using the **read_checkpoint** command.

    ```
    read_checkpoint -cell system_i/rst_ps7_0_50M
    Synth/Static/system_rst_ps7_0_50M_0.dcp
    ```

- Load the IP checkpoint for the auto pc by using the **read_checkpoint** command.

    ```
    read_checkpoint -cell
    system_i/ps7_0_axi_periph/s00_couplers/auto_pc
    Synth/Static/system_auto_pc_0.dcp
    ```

- Load the IP checkpoint for the system bus xbar by using the **read_checkpoint** command.

    ```
    read_checkpoint -cell system_i/ps7_0_axi_periph/xbar
    Synth/Static/system_xbar_0.dcp
    ```

- Load the IP checkpoint for the MAB by using the **read_checkpoint** command.

    ```
    read_checkpoint -cell design_1_i/MAB1
    Synth/Static/design_1_streamBlank_1_0.dcp

    read_checkpoint -cell design_1_i/MAB2
    Synth/Static/design_1_streamBlank_2_0.dcp

    read_checkpoint -cell design_1_i/MAB3
    Synth/Static/design_1_streamBlank_0_0.dcp
    ```

- Load the IP checkpoint for the Transfer IP by using the **read_checkpoint** command.

    ```
    read_checkpoint -cell design_1_i/machine1
    Synth/Static/design_1_informTransfer_0_0.dcp

    read_checkpoint -cell design_1_i/machine2
    Synth/Static/design_1_informTransfer_1_0.dcp

    read_checkpoint -cell design_1_i/machine3
    Synth/Static/design_1_informTransfer_2_0.dcp
    ```

- Load the IP checkpoint for the Compare by using the **read_checkpoint** command.

    ```
    read_checkpoint -cell design_1_i/Compare
    Synth/Static/design_1_compare_0_0.dcp

    read_checkpoint -cell design_1_i/Compare2
    Synth/Static/design_1_comparePR_0_0.dcp
    ```

- Load the IP checkpoint for the OFDM Transceiver by using the **read_checkpoint** command.

    ```
    read_checkpoint -cell design_1_i/OFDM
    Synth/Static/design_1_DPR_QAM_QPSK_0_0.dcp
    ```

**2-1-3.** Load one RM (UCB) for the MAB RP & two RMs (QAM modulation & QAM demodulation) by using the **read_checkpoint** command.

```
read_checkpoint -cell design_1_i/MAB1/inst/u1 Synth/UCB/ucb_synth.dcp

read_checkpoint -cell design_1_i/MAB2/inst/u1 Synth/UCB/ucb_synth.dcp

read_checkpoint -cell design_1_i/MAB3/inst/u1 Synth/UCB/ucb_synth.dcp

read_checkpoint -cell
design_1_i/OFDM/inst/DPR_QAM_QPSK_v1_0_S00_AXI_inst/T1/UUT/DAT_Mod_Ins
Synth/QAM/mod_synth.dcp

read_checkpoint -cell
design_1_i/OFDM/inst/DPR_QAM_QPSK_v1_0_S00_AXI_inst/T1/UUT1/DataSymDem_
ins Synth/QAM/demod_synth.dcp
```

## Define Reconfigurable Properties on each RM              Step 3

**3-1. In this design you have five Reconfigurable Partitions. Define the reconfigurable properties to the loaded RMs.**

**3-1-1.** Define each of the loaded RMs (submodules) as partially reconfigurable by setting the **HD.RECONFIGURABLE** property using the following commands.

```
set_property HD.RECONFIGURABLE 1 [get_cells
design_1_i/OFDM/inst/DPR_QAM_QPSK_v1_0_S00_AXI_inst/T1/UUT/DAT_Mod_Ins]

set_property HD.RECONFIGURABLE 1 [get_cells
design_1_i/OFDM/inst/DPR_QAM_QPSK_v1_0_S00_AXI_inst/T1/UUT1/DataSymDem_
ins]

set_property HD.RECONFIGURABLE 1 [get_cells design_1_i/MAB1/inst/u1]

set_property HD.RECONFIGURABLE 1 [get_cells design_1_i/MAB2/inst/u1]

set_property HD.RECONFIGURABLE 1 [get_cells design_1_i/MAB3/inst/u1]
```

This is the point at which the Partial Reconfiguration license is checked.

## Define the Reconfigurable Partition Region

**3-2. Next you must floorplan the RP regions. Depending on the type and amount of resources used by all the RMs for the given RP, the RP region must be appropriately defined so it can accommodate any RM variant.**

**3-2-1.** You execute the following command to define the region for each RP, perform the DRC.

```
read_xdc fplan.xdc
```

## Create and Implement First Configuration

**4-1. Create and implement the first Configuration.**

**4-1-1.** Execute the following command to implement the first configuration, the UCB algorithm inside the MAB block with QAM modulation scheme for the transceiver.

```
source create_first_configuration.tcl
```

The script will do the following tasks:

- The script will optimize, place and route the design by executing the following commands.

    ```
    opt_design

    place_design

    route_design
    ```

- Save the full design checkpoint.

    ```
    write_checkpoint -force Implement/UCB_QAM/top_ucb_qam_synth.dcp
    ```

    At this point, a fully implemented partial reconfiguration design from which full and partial bitstreams can be generated is ready. The static portion of this configuration **must** be used for all subsequent configurations, and to isolate the static design, the current reconfigurable module must be removed.

**4-2. After the first configuration is created, the static logic implementation will be reused for the rest of the configurations. So it should be saved. But before you save it, the loaded RM should be removed.**

**4-2-1.** Execute the following command to update the design with the blackbox and write the checkpoint.

```
source lock_placement_with_blackbox.tcl
```

The script will do the following tasks:

- Clear out the existing RMs executing the following commands.

    ```
    update_design -cell design_1_i/MAB1/inst/u1 -black_box

    update_design -cell design_1_i/MAB2/inst/u1 -black_box

    update_design -cell design_1_i/MAB3/inst/u1 -black_box

    update_design -cell design_1_i/OFDM/inst/DPR_QAM_QPSK_v1_0_S00_AXI_inst/T1/UUT/DAT_Mod_Ins -black_box

    update_design -cell design_1_i/OFDM/inst/DPR_QAM_QPSK_v1_0_S00_AXI_inst/T1/UUT1/DataSymDem_ins -black_box
    ```

    Issuing this command will result in design changes including, the number of Fully Routed nets (green) decreased, the number of Partially Routed nets (yellow) has increased, and *RPs* may appear in the Netlist view as empty.

- Lock down all placement and routing by executing the following command.

    ```
    lock_design -level routing
    ```

    Because no cell was identified in the `lock_design` command, the entire design in memory (currently consisting of the static design with black boxes) is affected.

- Write out the remaining static-only checkpoint by executing the following command.

```
write_checkpoint -force Checkpoint/static_route_design.dcp
```

This static-only checkpoint would be used for any future configuration, but here, you simply keep this design open in memory.

- Close the project.

```
Close_project
```

## Create Other Configurations

### 5-1.    Read next set of RM dcp, create and implement the second configuration.

**5-1-1.** Execute the following command to create and implement the second configuration, the UCB_T algorithm inside the MAB block with QPSK modulation scheme for the transceiver.

```
source create_second_configuration.tcl
```

The script will do the following tasks:

- First, it will open the blanking configuration using the tcl command:

```
open_checkpoint Checkpoint/static_route_design
```

- With the locked static design open in memory, read in post-synthesis checkpoint for the other reconfigurable modules.

```
read_checkpoint -cell design_1_i/MAB1/inst/u1
Synth/UCB_T/ucb_synth.dcp

read_checkpoint -cell design_1_i/MAB2/inst/u1
Synth/UCB_T/ucb_synth.dcp

read_checkpoint -cell design_1_i/MAB3/inst/u1
Synth/UCB_T/ucb_synth.dcp

read_checkpoint -cell
design_1_i/OFDM/inst/DPR_QAM_QPSK_v1_0_S00_AXI_inst/T1/UUT/DAT_Mo
d_Ins Synth/QPSK/mod_synth.dcp

read_checkpoint -cell
design_1_i/OFDM/inst/DPR_QAM_QPSK_v1_0_S00_AXI_inst/T1/UUT1/DataS
ymDem_ins Synth/QPSK/demod_synth.dcp
```

Optimize, place and route the design by executing the following commands.

```
opt_design

place_design

route_design
```

- Save the full design checkpoint.

```
write_checkpoint -force
Implement/UCBT_QPSK/top_ucbt_qpsk_synth.dcp
```

- Close the project

    ```
    close_project
    ```

# Create Other Configurations

## 6-1. Read next set of RM dcp, create and implement the third configuration.

**6-1-1.** Execute the following command to create and implement the third configuration, the UCB_V algorithm inside the MAB block with QPSK modulation scheme for the transceiver.

**6-1-2.** `source create_third_configuration.tcl`

The script will do the following tasks:

- First, it will open the blanking configuration using the tcl command:

    ```
    open_checkpoint Checkpoint/static_route_design
    ```

- With the locked static design open in memory, read in post-synthesis checkpoint for the second reconfigurable module.

    ```
    read_checkpoint -cell design_1_i/MAB1/inst/u1 Synth/UCB_V/ucb_synth.dcp

    read_checkpoint -cell design_1_i/MAB2/inst/u1 Synth/UCB_V/ucb_synth.dcp

    read_checkpoint -cell design_1_i/MAB3/inst/u1 Synth/UCB_V/ucb_synth.dcp

    read_checkpoint -cell design_1_i/OFDM/inst/DPR_QAM_QPSK_v1_0_S00_AXI_inst/T1/UUT/DAT_Mod_Ins Synth/QPSK/mod_synth.dcp

    read_checkpoint -cell design_1_i/OFDM/inst/DPR_QAM_QPSK_v1_0_S00_AXI_inst/T1/UUT1/DataSymDem_ins Synth/QPSK/demod_synth.dcp
    ```

- Optimize, place and route the design by executing the following commands.

    ```
    opt_design
    place_design
    route_design
    ```

- Save the full design checkpoint.

    ```
    write_checkpoint -force Implement/UCBV_QPSK/top_route_design.dcp
    ```

- Close the project

    ```
    close_project
    ```

# Create the blanking configuration.

**7-1-1.** Execute the following command to create and implement the second configuration

```
source create_blanking_configuration.tcl
```

The script will do the following tasks:

- Open the static route checkpoint.

  ```
  open_checkpoint Checkpoint/static_route_design.dcp
  ```

- For creating the blanking configuration, use the `update_design -buffer_ports` command to insert LUTs tied to constants to ensure the outputs of the reconfigurable partition are not left floating.

  ```
  update_design -buffer_ports -cell design_1_i/MAB1/inst/u1

  update_design -buffer_ports -cell design_1_i/MAB2/inst/u1

  update_design -buffer_ports -cell design_1_i/MAB3/inst/u1

  update_design -buffer_ports -cell
  design_1_i/OFDM/inst/DPR_QAM_QPSK_v1_0_S00_AXI_inst/T1/UUT/DAT_Mo
  d_Ins

  update_design -buffer_ports -cell
  design_1_i/OFDM/inst/DPR_QAM_QPSK_v1_0_S00_AXI_inst/T1/UUT1/DataS
  ymDem_ins
  ```

- Now place and route the design. There is no need to optimize the design.

  ```
  place_design

  route_design
  ```

  The base (or blanking) configuration bitstream, when we generate in the next section, will have no logic for either reconfigurable partition, simply outputs driven by ground. Outputs can be tied to VCC if desired, using the HD.PARTPIN_TIEOFF property.

- Save the checkpoint in the `BLANK` directory.

  ```
  write_checkpoint -force Implement/BLANK/top_ucb_synth.dcp
  ```

- Close the project

  ```
  Close_project
  ```

# Generate Bit Files

## 8-1. After all the Configurations have been validated by PR_Verify, full and partial bit files must be generated for the entire project

**8-1-1.** Generate the full configurations and partial bitstreams by executing the following tcl script.

```
source generate_bitstreams.tcl
```

**8-1-2.** The script will do the following tasks:

- Read the first configuration in the memory, using the command:

    ```
    open_checkpoint Implement/UCB_QAM/top_ucb_qam_synth.dcp
    ```

- Generate the full and partial bitstreams for this design.

    ```
    write_bitstream -file Bitstreams/UCB_QAM.bit

    close_project
    ```

- Read the blanking configuration in the memory, using the command:

    ```
    open_checkpoint Implement/BLANK/top_ucb_synth.dcp
    ```

- Generate a full bitstream with black boxes, plus blanking bitstreams for the reconfigurable modules. Blanking bitstreams can be used to "erase" an existing configuration to reduce power consumption.

    ```
    write_bitstream -file Bitstreams/BLANK.bit

    close_project
    ```

- Read the second configuration in the memory, using the command:

    ```
    open_checkpoint Implement/UCBT_QPSK/top_ucbt_qpsk_synth.dcp
    ```

- Generate full and partial bitstreams for the second configuration.

    ```
    write_bitstream -file Bitstreams/TUNED_QPSK.bit

    close_project
    ```

- Read the third configuration in the memory, using the command:

    ```
    open_checkpoint Implement/UCBV_QPSK/top_ucbt_qpsk_synth.dcp
    ```

- Generate full and partial bitstreams for the second configuration.

    ```
    write_bitstream -file Bitstreams/VAR_QPSK.bit

    close_project
    ```

## Generate the Software Application                                    Step 10

### 9-1. Open the PS design that was created in Step 1. Export the hardware design and launch SDK.

**9-1-1.** Click on the **Open Project** link, browse to *c:/Summer/Tutorial/tutorial*, select the *tutorial.xpr* and click **OK** to open the design created in Step 1.

**9-1-2.** Select **File > Export > Export Hardware…**

**9-1-3.** In the *Export Hardware* form, do not check the *Include bitstream* checkbox and click **OK**.

**9-1-4.** Select **File > Launch SDK**

**9-1-5.** Click **OK** to launch SDK.

The SDK program will open. Close the Welcome tab if it opens.

## 9-2. Create a Board Support Package enabling generic FAT file system library.

**9-2-1.** In **SDK**, select **File** > **New** > **Board Support Package.**

**9-2-2.** Click **Finish** with the default settings (with standalone operating system).

This will open the Software Platform Settings form showing the OS and libraries selections.

**9-2-3.** Select **xilffs** as the FAT file support is necessary to read the partial bit files.

| Name | Version | Description |
| --- | --- | --- |
| libmetal | 1.0 | Libmetal Library |
| lwip141 | 1.6 | lwIP TCP/IP Stack library: lwIP v1.4.1 |
| openamp | 1.1 | OpenAmp Library |
| ✓ xilffs | 3.4 | Generic Fat File System Library |
| xilflash | 4.2 | Xilinx Flash library for Intel/AMD CFI com... |
| xilisf | 5.7 | Xilinx In-system and Serial Flash Library |
| xilmfs | 2.1 | Xilinx Memory File System |
| xilpm | 2.0 | Power Management API Library for Zynq... |
| xilrsa | 1.2 | Xilinx RSA Library |
| xilskey | 6.0 | Xilinx Secure Key Library |

**Figure 13. Selecting the xilffs library support**

**9-2-4.** Click **OK** to accept the settings and create the BSP.

## 9-3. Create an application.

**9-3-1.** Select **File** > **New** > **Application Project.**

**9-3-2.** Enter **TestApp** as the *Project Name*, and for *Board Support Package*, choose **Use Existing** (*standalone_bsp_0* should be the only option).

**9-3-3.** Click **Next**, and select *Empty Application* and click **Finish.**

**9-3-4.** Expand the **TestApp** entry in the project view, right-click the *src* folder, and select **Import.**

**9-3-5.** Expand **General** category and double-click on **File System.**

**9-3-6.** Browse to *c:\Summer\Tutorial\Sources* and click **OK.**

**9-3-7.** Select **TestApp.c** and click **Finish** to add the file to the project.

**9-3-8.** Right-click on **TestApp** and select **C/C++ Building Settings.**

## 9-4. Create a zynq_fsbl application.

**9-4-1.** Select **File** > **New** > **Application Project.**

**9-4-2.** Enter **zynq_fsbl** as the *Project Name*, and for *Board Support Package*, choose **Create New**.

**9-4-3.** Click **Next**, select *Zynq FSBL,* and click **Finish.**

This will create the first stage bootloader application called zynq_fsbl.elf

## 9-5. Create a Zynq boot image.

**9-5-1.** Select **Xilinx Tools** > **Create Boot Image.**

**9-5-2.** Click the Browse button of the Output BIF file path field, browse to *c:\Summer\Tutorial,* and then click **Save** with the *output* as the default filename.

**9-5-3.** Click on the **Add** button of the *Boot image partitions,* click the Browse button in the Add Partition form, browse to *c:\Summer\Tutorial \tutorial\tutorial.sdk\zynq_fsbl\Debug* directory, select *zynq_fsbl.elf* and click **Open**.

**9-5-4.** Click **OK**. Click again on the **Add** button of the *Boot Image partitions,* click the Browse button in the Add Partition form, browse to *c:\Summer\Tutorial \Bitstreams* directory, select *BLANK.bit* and click **Open**.

**9-5-5.** Click **OK**.

**9-5-6.** Click again on the **Add** button of the *Boot Image partitions,* click the Browse button in the Add Partition form, browse to *c:\Summer\Tutorial \tutorial\tutorial.sdk\TestApp\Debug* directory, select *TestApp.elf* and click **Open**.

**9-5-7.** Click **OK**.

**9-5-8.** Make sure that the output path is *c:\Summer\Tutorial* and the filename is *BOOT.bin,* and click **Create Image**.

**9-5-9.** Close the SDK program by selecting **File > Exit**.

# Test the Design                                                                 Step 10

**10-1. Connect the board with micro-USB cable connected to the UART. Place the board in the SD boot mode. Copy the generated BOOT.bin and the partial bit files on the SD card and place the SD card in the board. Power On the board.**

**10-1-1.** Make sure that a micro-usb cable is connected to the UART port.

**10-1-2.** Make sure that the board is set to boot in SD card boot mode.

**10-1-3.** Using the Windows Explorer, copy the **BOOT.bin** and other partial binaries on to a SD Card.

**10-1-4.** Place the SD Card in the board and power ON the board.

**10-2. Start a terminal emulator program such as TeraTerm or HyperTerminal. Select an appropriate COM port (you can find the correct COM number using the Control Panel). Set the COM port for 115200 baud rate communication.**

**10-2-1.** Start a terminal emulator program such as TeraTerm or HyperTerminal.

**10-2-2.** Select the appropriate COM port (you can find the correct COM number using the Control Panel).

**10-2-3.** Set the *COM* port for **115200** baud rate communication.

**10-2-4.** Press BTN7 to display a menu.

**10-2-5.** Follow the menu and test various reconfigurations.

**10-2-6.** Below is an example user test to show how the terminal window will appear after various reconfigurations.

```
a. Algorithm Change     : 1<T>, 2<U>, 3<B>
b. Add a channel        : 4
c. Remove a channel     : 5
d. Run an experiment    : 6

User Input > 3
-> UCB configured.

a. Algorithm Change     : 1<T>, 2<U>, 3<B>
b. Add a channel        : 4
c. Remove a channel     : 5
d. Run an experiment    : 6

User Input > 6

**************************************************************
Running experiment for 10000 time slots.
Channel means are 0.43,0.92,0.35,0.87,0.41
Channel variances are 0.04,0.08,0.10,0.05,0.04

Switching from QPSK to QAM-16 for channel 2 at slot 1001
Switching from QPSK to QAM-16 for channel 4 at slot 1008

Channels 1,2,3,4,5 were selected 28,7287,27,2632,26 times.
**************************************************************

a. Algorithm Change     : 1<T>, 2<U>, 3<B>
b. Add a channel        : 4
c. Remove a channel     : 5
d. Run an experiment    : 6

User Input > 5
-> Channel removed.

a. Algorithm Change     : 1<T>, 2<U>, 3<B>
b. Add a channel        : 4
c. Remove a channel     : 5
d. Run an experiment    : 6

User Input > 6

**************************************************************
Running experiment for 10000 time slots.
Channel means are 0.43,0.92,0.35,0.87
Channel variances are 0.04,0.08,0.10,0.05

Switching from QPSK to QAM-16 for channel 2 at slot 1001
Switching from QPSK to QAM-16 for channel 4 at slot 1013

Channels 1,2,3,4 were selected 28,7772,26,2174 times.
**************************************************************
```



\end{document}